\documentclass[usegraphicx,usenatbib,useAMS]{mn2e}
\newcommand\ion[2]{#1{\thinspace\scshape#2}}%

\DeclareMathSymbol{\upalpha}{0}{UPM}{"0B}
\DeclareMathSymbol{\upbeta}{0}{UPM}{"0C}

\defcitealias{MH2007}{MH07}

\title[Probing Reionization with Quasar Spectra]{Probing Reionization
  with Quasar Spectra: the Impact of the Intrinsic Lyman-$\upalpha$
  Emission Line Shape Uncertainty} 

\author[R. H. Kramer and Z. Haiman]{R. H. Kramer$^{1}$\thanks{E-mail:
    roban@astro.columbia.edu (RHK); zoltan@astro.columbia.edu (ZH)}
  and Z. Haiman $^{1}$\footnotemark[1]\\ $^{1}$Department of
  Astronomy, Columbia University, 550 West 120th Street, New York, NY
  10027}

\begin{document}

\date{}

\pubyear{2009}

\maketitle

\begin{abstract}
Arguably the best hope of understanding the tail end of the
reionization of the intergalactic medium (IGM) at redshift $z > 6$ is
through the detection and characterization of the Gunn-Peterson (GP)
damping wing absorption of the IGM in bright quasar spectra. However,
the use of quasar spectra to measure the IGM damping wing requires a
model of the quasar's intrinsic Lyman-$\upalpha$ emission line. Here
we quantify the uncertainties in the intrinsic line shapes, and how
those uncertainties affect the determination of the IGM neutral
fraction. We have assembled a catalog of high-resolution HST spectra
of the emission lines of unobscured low-redshift quasars, and have
characterized the variance in the shapes of their lines. We then add
simulated absorption from the high-redshift IGM to these quasar
spectra in order to determine the corresponding uncertainties in
reionization constraints using current and future samples of $z > 6$
quasar spectra. We find that, if the redshift of the Lyman-$\upalpha$
emission line is presumed to coincide with the systemic redshift
determined from metal lines, the inferred IGM neutral fraction is
systematically biased to low values due to a systematic blueshift of
the Lyman-$\upalpha$ line relative to the metal lines. If a similar
blueshift persists in quasars at $z>6$, this bias strengthens previous
claims of a significant neutral hydrogen fraction at $z \approx 6$.
The bias can be reduced by including a Lyman-$\upalpha$ blueshift in
the modeling procedure, or by excising wavelengths near the
Lyman-$\upalpha$ line center from the modeling. Intrinsic
Lyman-$\upalpha$ line shape variations still induce significant
scatter in the inferred $x_\mathrm{IGM}$ values.  Nevertheless, this
scatter still allows a robust distinction between a highly ionized
($x_\mathrm{IGM} \sim 10^{-3}$) and a neutral ($x_\mathrm{IGM} = 1$)
IGM with even a few bright quasars.  We conclude that if the
variations of the intrinsic Lyman-$\upalpha$ emission line shapes in
high-$z$ quasars are similar to those at low-$z$, this variation will
not limit the usefulness of quasar spectra in probing reionization.
\end{abstract}


\section{Introduction}

One of the most important frontiers of observational astronomy is the
study of the reionization of the universe. The details of reionization
--- when it started, how long it took, and what types of objects
contributed ionizing photons --- hold clues to the birth and early
history of the various structures we see in the universe
today. Because of their high luminosities and relatively flat spectra,
quasars are excellent probes of reionization. Many efforts have been
made to constrain the ionization state of the intergalactic medium
(IGM), and thus the timing of the tail-end of reionization, using
spectra of $z > 6$ quasars discovered by the Sloan Digital Sky Survey
\citep*[][and references therein]{FCK2006}. Beginning at lower
redshift, the forest of Lyman-$\upalpha$ absorption lines in a quasar
spectrum traces the fluctuating neutral hydrogen density along the
line of sight. At $z \sim 6$, the Lyman-$\upalpha$ forest lines blend
together and saturate, forming the dark Gunn-Peterson trough
\citep{GunnPeterson1965}, indicating a neutral fraction of $\ga
10^{-3}$\citep{Fan2006}. This saturation at such a small neutral
fraction limits the utility of Lyman-series absorption spectra as
direct probes of the reionization epoch, though the scattered
transmission windows in this otherwise dark trough do contain some
information about the IGM ionization state \citep{FCK2006}.

Fortunately, absorption is not confined to photons in the narrow,
resonant core of the line. Photons passing through the neutral
hydrogen in the IGM can be absorbed in the damping wings of the
Lyman-$\upalpha$ line, far from the central resonance wavelength. This
means that the hydrogen that creates the Gunn-Peterson (GP) trough
also has a damping wing extending redward of the edge of the trough
\citep{Miralda-Escude1998}. If the neutral hydrogen extends all the
way up to the redshift of the source, then the GP trough will extend
up to the wavelength of the source's Lyman-$\upalpha$ emission line. A
quasar, however, is an extremely luminous source of ionizing
radiation. This radiation ionizes a bubble of the surrounding IGM. The
bubble is large enough (tens of comoving Mpc) to extend over a
significant redshift interval, which pushes the edge of the GP trough
to lower redshift, opening up a transmission window in the quasar
spectrum blueward of its central Lyman-$\upalpha$ emission
wavelength. The damping wing of the GP trough extends into this
window, where its characteristic absorption profile can be detected
and its magnitude used to constrain the IGM neutral fraction. Since
the damping wing is many orders of magnitude weaker than absorption in
the core of the line, this technique can probe much higher neutral
hydrogen fractions without saturation \citep{CH2000, MadauRess2000}.

\citet[][hereafter \citetalias{MH2007}]{MH2007} searched for the
signature of the damping wing in spectra of three $z > 6.2$ quasars,
finding best-fit values for the IGM neutral fraction of
$x_\mathrm{IGM} = 1.0$, $1.0$, and $0.2$, at $z = 6.22$, $6.28$, and
$6.4$ respectively.\footnote{Throughout this paper we use
  $x_\mathrm{IGM}$ to mean the neutral fraction at mean IGM
  density. This can differ from both the volume-weighted neutral
  fraction and the mass-weighted neutral fraction, depending on the
  clumping factor and density distribution of the IGM.} The
uncertainty in this inferred value is large, but they determined a
lower limit of $x_\mathrm{IGM} > 0.04$ for the first two sources. In
an earlier analysis of the spectrum of the $z = 6.28$ quasar,
\citet{MH2004} determined $x_\mathrm{IGM} \ga 0.2$ using both the
Lyman-$\upalpha$ and Lyman-$\upbeta$ regions of the spectrum. Other
attempts to use the transmission window to measure $x_\mathrm{IGM}$
have focused on the size of the ionized region and found similar
results \citep{WyitheLoeb2004}. Fluctuations in the IGM density and
ionizing background (as well as uncertainties in assumptions about the
quasar age and luminosity) represent significant sources of
uncertainty in these estimates \citep{BH2007, Maselli2007}.

There are two main sources of uncertainty in determining the IGM
ionization state using the GP damping wing. The first is the
uncertainty in the underlying emission spectrum of the quasar. The
damping wing absorption profile is seen against the intrinsic spectrum
of the quasar, which must be modeled accurately in order to accurately
calculate the optical depth profile. The second source of uncertainty
is in modeling the IGM absorption profile itself. Both the damping
wing and the superimposed absorption from residual neutral hydrogen
inside the ionized region are affected by the density structure and
ionizing background in the IGM. These effects must be incorporated
into the model that is fit to the observed spectra.

In this paper we address only the uncertainty caused by errors in the
determination of the intrinsic quasar spectrum. Since the edge of the
Gunn-Peterson trough is at only a slightly lower redshift than the
quasar itself, the absorption profile we want to measure is
superimposed on the blue wing of the quasar's Lyman-$\upalpha$
emission line (nominally centered at $1215.67$~\AA\ in the rest
frame). Therefore, determination of the intrinsic flux requires that
we model the quasar's emission line profile accurately, and be able to
fit the model to the observed spectrum \textit{using only the
  unabsorbed red wing of the line profile}. Random errors in this
process will add to the \textit{scatter} in recovered values. More
worryingly, a bias in the flux modeling (a consistent asymmetry in the
profile that was not included in the model, for instance), would
\textit{bias} the $x_\mathrm{IGM}$ results.

The purpose of the present paper is to quantify both the bias and the
scatter in the inferred IGM neutral fraction (and other model
parameters) that arise from variations of the intrinsic
Lyman-$\upalpha$ emission line shape. More specifically, we seek to
answer the following questions: (i) how accurately do
\citetalias{MH2007} extrapolate the flux from the red side of the
Lyman-$\upalpha$ line to the blue side, (ii) how do errors in the
extrapolation affect the IGM neutral fraction determination, and (iii)
could flux errors bias the $x_\mathrm{IGM}$ value and cause an ionized
IGM to mimic a neutral one, invalidating their results?

The first step in answering these questions is to understand the
intrinsic shapes of quasar Lyman-$\upalpha$ emission
lines. High-redshift quasars obviously have too much absorption in the
blue wing of the line to be useful for this purpose. We found that
even at $z=2.1$, where the Lyman-$\upalpha$ line enters the Sloan
Digital Sky Survery (SDSS) spectral wavelength range, the
Lyman-$\upalpha$ forest is often thick enough to interfere with a
precision study of the line profile. Therefore we chose to study a
library of spectra observed in the ultra-violet (UV) by the Hubble
Space Telescope, primarily at $z<1$. This catalog of spectra, and the
fits we perform in \S \ref{fullFits}, may also be useful in the study
of quasar environments and the quasar ``proximity effect'' at $z <
6$. We discuss this possibility further in \S \ref{futureSec}.

The rest of this paper is structured as follows. In \S \ref{lowzSec},
we discuss the library of low-$z$ spectra that we assembled. The
results of the line profile fits are discussed in \S
\ref{fullFits}. In \S \ref{halfFits} we analyze our ability to predict
the flux on the blue side of the line using only the flux on the red
side accurately at high-redshift. We describe our simulations of
high-redshift absorption spectra in \S \ref{simulatedSec}, and the impact
of extrapolation errors on the IGM neutral fraction recovery in \S
\ref{simulatedResSec}. In \S \ref{conclusionSec} we offer our
conclusions. Finally, in \S \ref{futureSec}, we discuss the future
potential of the techniques outlined here.

\section{Studying low-$z$ quasar Lyman-$\upalpha$ emission line profiles}\label{lowzSec}

\subsection{Assembling a library of quasar spectra}\label{sampleSec}

Our sample of low-redshift, unobscured quasars was selected from the
available HST archival data in a multi-stage process. Our criteria for
selection included the instrument used, spectral resolution,
wavelength coverage, and signal-to-noise ratio. Each entry in the HST
data archive (part of the Multimission Archive at the Space Telescope
Science Institute,
MAST\footnote{{http://archive.stsci.edu/hst/search.php}}) is tagged
with ``a short description of the target, supplied by the
observer''\footnote{{http://archive.stsci.edu/hst/help/search\_help.html}}. We
search for all Faint Object Spectrograph (FOS) and Goddard
High-Resolution Spectrograph (GHRS), observations of objects with
descriptions including ``QSO'' or ``QUASAR''. This yielded 2742
datasets. We found 856 unique RA and Dec coordinate pairs in this
list, making no attempt to eliminate close matches at this stage.

\begin{figure}
  \includegraphics[width=3in]{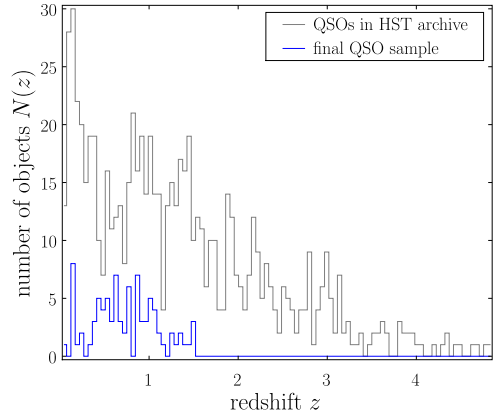}
  \caption{Redshift distribution of the sample. The upper (grey) line
    shows all of the NED redshifts we found for quasars in the MAST
    Hubble archive. The lower (blue) line is for the sample of 87
    spectra we adopted for this project (see \S
    \ref{sampleSec}).}\label{redshiftFig}
\end{figure}

In order to identify our objects and find their redshifts, we searched
the NASA Extragalactic Database (NED) for objects of type ``QSO'' near
each set of coordinates, taking the redshift of the closest match
($93\%$ were found). We next eliminated all observations that did not
cover the objects' redshifted Lyman-$\upalpha$ wavelength (calculated
from the NED redshift), which excluded about $44\%$ of the objects. We
found additional spectral datasets for these same objects by searching
MAST for observations at the coordinates of the already-identified
objects, then selecting additional spectra containing the
Lyman-$\upalpha$ line. We eliminated all observations that were not
taken with a grating (i.e. prism or mirror observations). At this
point we ran our code to align and coadd the selected spectra, as
described in Appendix \ref{coadding}. We then eliminated spectra that
did not include a designated wavelength range surrounding the line
($1200$--$1286$~\AA\ rest frame, cutting the object list by another
$15\%$), and excluded spectra with a mean signal-to-noise ratio of
less than 10 (eliminating $42\%$ of the remaining objects). This left
120 spectra of 112 objects (only one GHRS spectrum, the rest
FOS). After running our line-profile fits, we eliminated 16 more
spectra (14 objects) for which our line-profile fitting code failed to
converge (usually in the emission/absorption feature detection stage;
see Appendix \ref{featureDetection}). After visual inspection of the
full-profile fits we cut 21 more spectra (20 objects), 8 for poor data
quality (obvious bias in the flux over part of the wavelength range,
for instance), 2 because the spectra lacked visible emission features,
and 11 because of excessive absorption, either in the form of
Lyman-$\upalpha$ forest lines or broad absorption lines. This left 86
spectra (all FOS) of 78 objects as our final sample. Table
\ref{sampleTable} lists all 120 automatically-selected spectra, with
notes on the 21 spectra that were cut, as well as on other unusual
spectra or fits. Figure \ref{redshiftFig} shows the redshift
distributions of all of the quasars found in the HST archives, and of
those selected for this project. We have not, so far, used any Space
Telescope Imaging Spectrograph (STIS) data, but this would increase
our sample considerably (perhaps by a factor of two).

\begin{table*}
\begin{minipage}{6in}
\caption{HST (FOS and GHRS) quasar Lyman-$\upalpha$
  emission line spectra.}\label{sampleTable}
\begin{tabular}{p{1.1in}rcccrp{1.75in}}
\hline
Name & $z$ & Instrument & Aperture & Grating & S$/$N & Comments on spectra\\
(NED) & (NED) &  &  &  & (mean) & and fits.\\
\hline
  3C 273&0.158&FOS&B-2&H13&  19.9&                                                                                 \\

[HB89] 1427+480&0.221&FOS&B-3&H13&  12.8&               Absorption feature in blue wing of the line (flagged in full fit).\\
PG 0953+414&0.234&FOS&C-2&H13&  11.1&                                               Absorption feature at line center.\\

[HB89] 1156+213&0.349&FOS&B-3&H19&  10.7&                                                                                 \\
PG 1049-005&0.360&FOS&C-2&H19&  13.6&
Excluded due to bad blue edge of spectrum.\\
\hline
\end{tabular}
\medskip
The name and redshift $z$ are from NED. This is a sample of the full
table, which is available in the electronic version.
\end{minipage}
\end{table*}

\subsection{Defining a spectral model}

In order to characterize the uncertainties inherent in extrapolating a
high-redshift quasar's spectrum from the red side of the
Lyman-$\upalpha$ line to the blue side, we performed two kinds of fits
on our low-redshift spectra: one using the full profile visible in the
low-$z$ spectra, and one using only the data from the red side of the
line (mimicking a fit to a high-$z$ spectrum).

We follow \citetalias{MH2007} and adopt the commonly-used spectral
model consisting of a power-law continuum, a Gaussian \ion{N}{v} line,
and a double-Gaussian Lyman-$\upalpha$ line. Although the \ion{N}{v}
line is actually a doublet ($1238.821$ and $1242.804$
\AA)\footnote{http://physics.nist.gov/PhysRefData/ASD/lines\_form.html},
we fit it with a single Gaussian since the lines are quite broad and
thoroughly blended. On the other hand, the Lyman-$\upalpha$ line
usually has prominent broad and narrow components in quasar spectra
(thought to arise from gas close to, and relatively farther from the
central black hole, respectively), so we fit it with two independent
Gaussians. Note that the components found in our fits do not
necessarily correspond to physically distinct emission regions, but
this combination does yield good phenomenological fits to the line
profiles, which is our primary concern in this paper.

\begin{figure} 
    \includegraphics[width=3in]{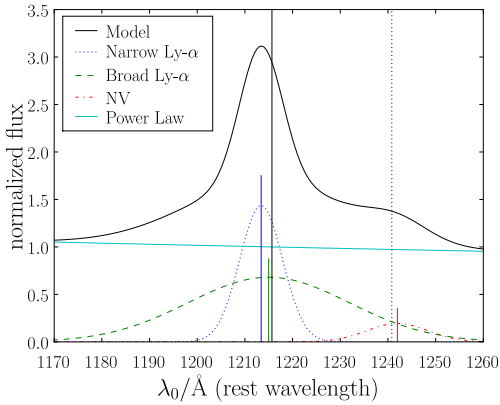}
    \caption{A model for a quasar spectrum (3C 273) in the vicinity of
      the Lyman-$\upalpha$ emission line. Components (as given in the
      legend) are: broad and narrow Lyman-$\upalpha$ Gaussians, a
      \ion{N}{v} Gaussian, and a power-law continuum. The centers of
      the Gaussian components (which are shifted significantly
      relative to their nominal wavelengths based on the NED redshift)
      are indicated with short vertical lines. Complete vertical lines
      indicate the nominal central wavelengths of Lyman-$\upalpha$
      (solid) and \ion{N}{v} (dotted). \label{lineModel}}
\end{figure}

Our spectral model, illustrated by the best-fit to a typical spectrum
(the quasar 3C 273) is shown in Figure \ref{lineModel}. The line
centers of the three Gaussians, especially that of the narrow
Lyman-$\upalpha$ component, are shifted significantly from their
nominal wavelengths (as discussed in \S \ref{shiftSec}). The model has
three parameters for each Gaussian (width, central wavelength,
amplitude) and two for the power-law (index and normalization), for a
total of 11 free parameters. In practice, the power-law is fit first,
then a separate fit is performed with the power-law parameters fixed
and the emission line parameters free. In some fits the power-law
index or line centers are also fixed (see \S \ref{fullFits} and
\ref{halfFits}).

\subsection{Procedure for line-profile fitting}\label{fitSec}

Each fit is performed using Levenberg-Marquardt $\chi^2$
minimization.\footnote{Fits use the \texttt{PDL::FIT::Levmar}
  package\\ (http://search.cpan.org/\~{ }jlapeyre/PDL-Fit-Levmar/)
  based on \texttt{levmar} (http://www.ics.forth.gr/\~{
  }lourakis/levmar/).} We perform our fits separately for the
continuum and the combined emission lines. First we fit the power-law
alone to regions of the spectrum chosen to be free of emission
features. The power-law fit regions are, in the rest frame: $[1155,
  1165]$, $[1280, 1293]$, $[1315, 1325]$, and $[1340, 1360]$~\AA. Some
spectra do not include this full wavelength range. In any fit to a
spectrum that lacks data at $\lambda_0 < 1165$~\AA\ we fixed
the power law index to $-1.3$, since we have found that extrapolating
the continuum from only the longer-wavelength regions can be
unreliable.

Once the index and normalization of the power-law are determined and
fixed, the emission line components are fit to a region centered on
the Lyman-$\upalpha$ line at $1215.67$ \AA, plus two neighboring
continuum-dominated regions. The continuum regions are included to
help constrain the widths of extremely broad Gaussian components. The
Gaussian fit regions are, in the rest frame: $[1155, 1165]$, $[1185,
  1250]$, and $[1280, 1293]$ \AA.

Initial values for the model parameters are given below. These were
chosen by trial and error to speed convergence, and are not
necessarily representative of the best-fit values. The powerlaw index
and normalization start at
\begin{equation}
p = -1.3; \\ a_\mathrm{p} = \tilde{F}_\mathrm{PL},
\end{equation}
where $\tilde{F}_\mathrm{PL}$ is the median flux in the power-law fit
region and the continuum flux is given by
\begin{equation}
  F_\lambda = a_\mathrm{PL} \left(\frac{\lambda}{\lambda_0}\right)^p,
\end{equation}
with $\lambda_0 = 1215.67(1+z)$~\AA.
The Lyman-$\upalpha$ narrow-component width, center, and amplitude
start at
\begin{equation}
\sigma_\mathrm{L\upalpha0} = 4~(1+z) ~\mathrm{\AA};\\
\mu_\mathrm{L\upalpha0} = 1215.67 ~\mathrm{\AA};\\
a_\mathrm{L\upalpha0} = \tilde{F}_\mathrm{G}.
\end{equation}
The Lyman-$\upalpha$ broad-component width, center, and amplitude
start at
\begin{equation}
\sigma_\mathrm{L\upalpha1} = 8~(1+z) ~\mathrm{\AA};\\
\mu_\mathrm{L\upalpha1} = 1215.67 ~\mathrm{\AA};\\
a_\mathrm{L\upalpha1} = \tilde{F}_\mathrm{G}.
\end{equation}
The \ion{N}{v}  width, center, and amplitude start at
\begin{equation}
\sigma_\mathrm{NV} = 2~(1+z) ~\mathrm{\AA};\\
\mu_\mathrm{NV} = 1240.81 ~\mathrm{\AA};\\
a_\mathrm{NV} = 2 \tilde{F}_\mathrm{G},
\end{equation}
where $z$ is the redshift of the quasar (from NED),
$\tilde{F}_\mathrm{G}$ is the median flux in the Gaussian fit region,
and the flux for Gaussian component $i$ is given by
\begin{equation}
F_{\lambda,i} = a_i \exp{\left[ \frac{- (\lambda - \mu_i)^2 }{ 2 \sigma_i^2}\right]}.
\end{equation}

While we chose to use low-redshift quasars in order to study
relatively unobscured spectra, some absorption features do appear from
both the Galactic interstellar medium (ISM) and the IGM. In order to
avoid ISM absorption lines, we exclude a small region around all of
the lines listed by \citet*{Verner1994}. The half-width of the excluded
region is given by $w = \sqrt{ w_\mathrm{min}^2 + w_\mathrm{Inst}^2}$,
where $w_\mathrm{min} = 1.5$~\AA\ and $w_\mathrm{Inst} =
2~\mathrm{FWHM} / 2.355$, with $\mathrm{FWHM}$ being the full width at
half maximum of the instrumental line spread function (LSF).

We use an iterative scheme described in Appendix
\ref{featureDetection} to exclude other absorption features. In order
to avoid biasing the results, we apply essentially the same exclusion
criteria to positive and negative deviations, meaning we exclude both
excess absorption and emission. In practice very few pixels are
flagged as emission features (except for some spectra in which broad
emission lines intrude into the power-law fit regions), while most
absorption features obvious to the eye (or in the ISM list) are
flagged, with few apparently spurious detections. Once this feature
detection scheme converges, we perform a final fit on the masked
spectrum to determine the best-fit parameters.

\section{The Shapes of Low-$z$ Quasar Line Profiles}\label{fullFits}
\subsection{Quality of the line-profile fits}

\begin{figure}
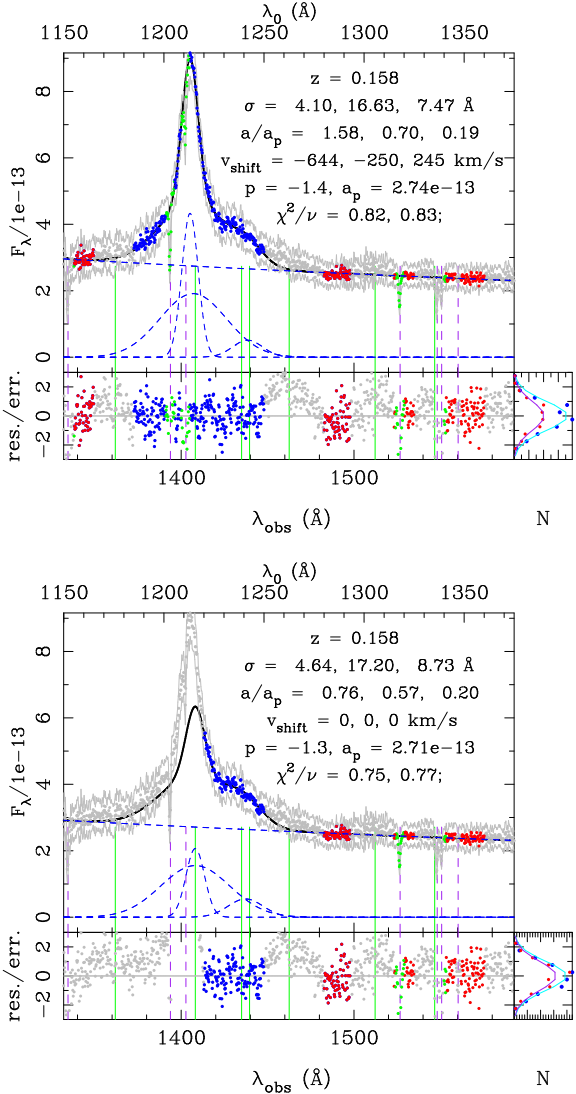

\includegraphics[height=3in,angle=270]{fig03a.eps}\\

\includegraphics[height=3in,angle=270]{fig03b.eps}
\caption{Spectral model fit to the observed spectrum of 3C~273 using
  the full line profile (top plot) and using only data redward of
  Lyman-$\upalpha$ (bottom plot). The points in the upper panels show
  the observed flux ($\mathrm{erg~s^{-1}~cm^{-2}~\AA^{-1}}$), with
  error envelope in grey. The best-fit model is the black line (often
  hidden by the data). Vertical solid (green) lines indicate expected
  broad emission lines (at the quasar redshift), while vertical dashed
  (purple) lines indicate expected narrow ISM absorption lines (at
  $z=0$). Red points were used for the power-law continuum fit. Blue
  points were used for the line-profile fit. Green points were
  excluded because of the presence of absorption lines. Grey points
  were not included in the fit. The top axis is labeled in the quasar
  rest frame, the bottom in the observer frame. The bottom panels show
  the residuals divided by the quoted formal measurement error. At the
  bottom-right are sideways histograms of the residuals (points) with
  the expected Normal distributions shown by the solid lines.  The
  text in the upper panels gives the redshift $z$; the widths $\sigma$
  of the narrow Ly-$\upalpha$, broad Ly-$\upalpha$, and \ion{N}{v}
  components; the height of each component relative to the continuum
  $a/a_\mathrm{p}$; the velocity shifts $v_\mathrm{shift}$ of each
  component; the powerlaw index $p$; the continuum amplitude at
  Ly-$\upalpha$ line center $a_\mathrm{p}$
  ($\mathrm{erg~s^{-1}~cm^{-2}~\AA^{-1}}$); and the reduced $\chi^2$
  values $\chi^2/\nu$ for the continuum and line-profile fits,
  respectively.
\label{specFitFig}}
\end{figure}

\begin{table}
\caption{Quasar Emission Lines Near Lyman-$\upalpha$}
\label{emissionLineTable}
\begin{tabular}{lr}
\hline\\
Line & Rest-Frame Wavelength\\
\ & $\lambda_0$ (\AA )\\ 
\hline\\
\ion{C}{iii}*    &  1175.70\\ 
Lyman-$\upalpha$ &  1215.67\\
\ion{N}{v}       &  1238.82\\
\ion{N}{v}       &  1242.80\\
\ion{Si}{ii}     &  1262.59\\
\ion{O}{i}+\ion{Si}{ii} &  1305.53\\
\ion{C}{ii}      &  1335.31\\
\ion{Si}{iv}+\ion{O}{iv}] & 1399.80\\
\hline\\
\end{tabular}

\medskip

\raggedright Wavelengths are from the SDSS Reference Line List
({http://www.sdss.org/dr6/algorithms/speclinefits.html}) with
\ion{Si}{ii} and \ion{C}{iii}* from \cite{SDSSComposite2001} and
\ion{N}{v} doublet from the Atomic Line List
({http://www.pa.uky.edu/\~{ }peter/atomic/}) compiled by Peter van
Hoof.

\end{table}

Before proceeding to our main goal of studying the impact of
emission-line shape variations of fits to high-$z$ quasars, in this
section we describe how well the full profiles can be fit with our
adopted spectral model. An example of a typical fit to a quasar
spectrum is shown in Figure \ref{specFitFig} (top plot). The upper
panel shows the spectrum, while the lower panel shows the
residuals. Pixels included in the fits are shown as colored points
(red for the continuum fits and blue for the line-profile fits). The
model spectrum (black solid line, almost entirely hidden by data
points) is clearly a good fit to the data used in the fits, as can be
seen visually and by the small reduced $\chi^2$ values. Vertical solid
lines (green) indicate the nominal rest wavelengths (listed in Table
\ref{emissionLineTable}) of broad quasar emission lines. Vertical
dashed lines indicate ISM absorption features listed in
\citet{Verner1994}. Some of these lines are also visible in the
residuals, and these and other pixels excluded from the fits
automatically by our absorption-feature detection code are indicated
in purple. The bottom-right panel shows sideways histograms of the
residuals (blue points for the line-profile fit, red for the continuum
fit) with the expected Normal distributions shown by the solid lines,
confirming that the distribution of the residuals matches the expected
distribution fairly well. The best-fit values of the model parameters
are printed on the graph. Table \ref{fullparamTable} lists the
best-fit model parameters for all of the full-profile fits. The
amplitudes of the Gaussian components are normalized to the continuum
flux at Lyman-$\upalpha$ (which is stated in the flux units of the
spectra, $\mathrm{erg~s^{-1}~cm^{-2}~\AA^{-1}}$). Also given are the
reduced $\chi^2$ values and the numbers of degrees of freedom for the
continuum and line-profile fits.

\begin{table*}
\caption{Best-fit spectral parameters from fits to the full line
  profile.}\label{fullparamTable}
\medskip
\hrule
\bigskip
Complete table available in the electronic version.
\medskip
\hrule
\bigskip

The first two columns give the NED name of the target, plus the HST
instrument, grating, and aperture used to take the
spectrum. $\sigma_\mathrm{L\upalpha0}$ is the width of the narrow
Lyman-$\upalpha$ component (corrected for redshift using the NED value
of $z$, see Table 1). $a_\mathrm{L\upalpha0}$ is the amplitude of this
component, normalized by the continuum flux
$a_\mathrm{p}$. $v_\mathrm{L\upalpha0}$ is the shift of the component
center from the nominal wavelength calculated with the NED
redshift. The next six columns give the same quantities for the broad
Lyman-$\upalpha$ component ($\mathrm{L\upalpha1}$), and the \ion{N}{v}
component. $p$ is the continuum power law index, and $a_\mathrm{p}$ is
the continuum flux at the nominal Lyman-$\upalpha$ central wavelength,
in units of $10^{-15} \mathrm{erg~s^{-1}~cm^{-2}~\AA^{-1}}$. The last
columns give the reduced $\chi^2$ value $\chi^2/\nu$ and the number of
degrees of freedom $\nu$ for the continuum and line-profile fits.

\end{table*}

\begin{figure}
 \includegraphics[width=3in]{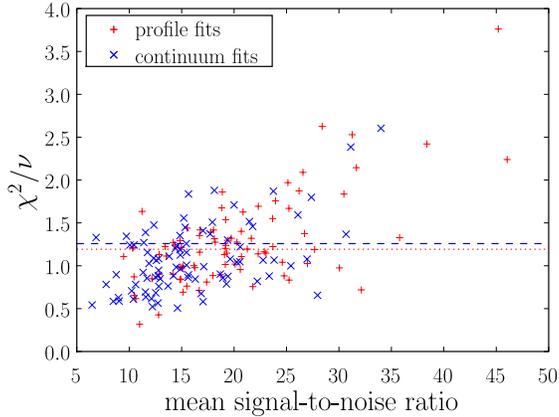}
 \caption{Reduced $\chi^2$ values versus mean signal-to-noise ratio
   for the line profile ($+$) and continuum (x) fits. Both values are
   calculated using only the pixels included in each fit. The
   $\chi^2/\nu$ values corresponding to the $99\%$ significance level
   (using the median number of degrees of freedom: 183 for continuum
   and 318 for profile fits) are indicated in the figure by horizontal
   lines (at $\chi^2/\nu = 1.26$ and $1.19$ respectively).
\label{snFig}}
\end{figure}

Figure \ref{snFig} plots the reduced $\chi^2$ value, $\chi^2/\nu$, for
each of our fits versus the mean signal-to-noise ratio for the pixels
used in that fit. A large fraction of the fits (41/87 profile and
24/87) are formally unacceptable at the $99\%$ significance level. The
$\chi^2/\nu$ values corresponding to this significance level (using
the median number of degrees of freedom: 183 for continuum and 318 for
profile fits) are indicated in the figure by horizontal lines (at
$\chi^2/\nu = 1.26$ and $1.19$ respectively). To the eye, our models
generally look like good matches to the overall shape of the line
profile. Examination of the residuals suggests that the formally poor
fits are due primarily to small deviations at the $\sim 5$--$10\%$
level. This is more than adequate performance for our purposes, but it
is interesting to note that the intrinsic quasar spectra are, at some
level, more complicated than our models.

\subsection{Shifts in Emission Components}\label{shiftSec}

\begin{figure}
 \includegraphics[width=3in]{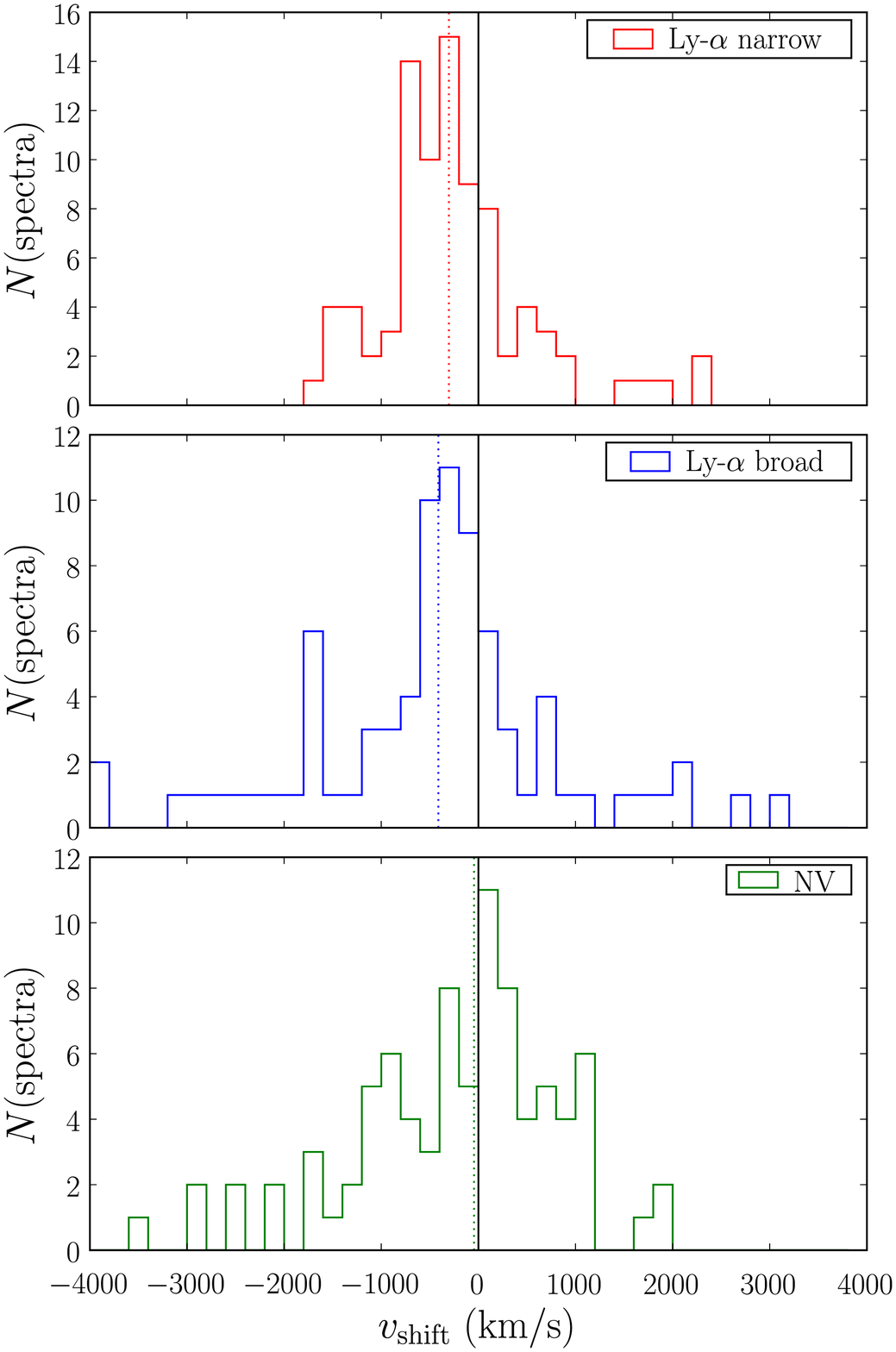}
 \caption{Distribution of the shifts of three emission-line components
   relative to the systemic redshift. Median values are indicated by
   vertical dotted lines. The median shifts are: $-303~\mathrm{km/s}$
   for the narrow component, $-413~\mathrm{km/s}$ for the broad
   component, and $-45~\mathrm{km/s}$ for the \ion{N}{v} component. A
   handful of extreme values fell outside of the plotted
   range. \label{vshiftFig}}
\end{figure}

Since our primary concern in this paper is the effect of errors in the
extrapolation of intrinsic quasar spectra from the red side of the
line to the blue side of the line, we pay particular attention to any
shifts or asymmetries in the line profiles that could introduce a
systematic bias in the extrapolation. Obviously it is easiest
to extrapolate the line profiles from the red side to the blue side if
they are symmetric and consistently centered at a well-defined
wavelength. Figure \ref{vshiftFig} shows the distributions of shifts
of emission component centers from the fits to all of the spectra. The
shifts in the line center wavelength $\mu$ relative to the nominal
central wavelength $(1+z)\lambda_{0}$ has been represented as a
velocity
\begin{equation}
v_\mathrm{shift} = c~\frac{\mu -
  (1+z)\lambda_{0}}{(1+z)\lambda_{0}}
\end{equation}
where $z$ is the NED redshift. These shifts are often hundreds or even
thousands of $\mathrm{km/s}$, and show little correlation between
components. Caution must be exercised in interpreting these velocity
values, since our line-profile model is phenomenological rather than
physical. Nonetheless, many line profiles clearly show distinct
emission components shifted by large velocities relative to their
nominal wavelengths and to each other.

The nominal central wavelengths are calculated using the redshifts
from NED, which were mostly determined from low-ionization metal
lines. It is well known that there is often a velocity offset between
high-ionization lines and low-ionization lines in quasar spectra
\citep[see discussion and references in][]{Shang2007}. There have been
relatively few studies of shifts in the Lyman-$\upalpha$ and
\ion{N}{v} lines, in part because of the difficulty of separating the
components. \citet{Shang2007} measured velocity shifts of the peak of
the Lyman-$\upalpha$ line relative to the [\ion{O}{iii}]
$5006.8$~\AA\ line for a sample of nearby quasars ($z < 0.4$). This
should roughly correspond to our narrow component shifts, since this
component dominates in the peak of the overall profile (see
Fig. \ref{lineModel}). To check this we compared the shifts of line
profile peaks versus the best-fit narrow component shift in our sample
and found that (with very few exceptions) they indeed correspond
closely. Their sample has a mean shift of $-90~\mathrm{km/s}$ with
standard deviation of $250~\mathrm{km/s}$. They also measure the
asymmetry of the total line profile, finding that the Lyman-$\upalpha$
line almost always has an excess in the blue wing, indicating that the
broad component tends to be shifted blueward of the narrow
component. Their measurements of the shifts of other lines suggest
that lines arising from higher ionization states are shifted by larger
amounts. Of the lines they measured, they find that the
Lyman-$\upalpha$ shift is well correlated only with the \ion{C}{iv}
line at $1549.48$~\AA.

\citet{1999McIntosh} on the other hand, found a mean velocity shift of
$-550~\mathrm{km/s}$ between the Lyman-$\upalpha$ line and the
~[\ion{O}{iii}] lines for a sample of quasars at $2.0 \leq z \leq
2.5$. It is not clear why their distribution is so different, but it
may be due to sampling a different population of higher-luminosity
quasars (the luminosity limit was $V \leq 18$ versus $B < 23$).

There is no consensus on a physical model explaining these shifts, or
indeed on a physical model of quasar broad-line and narrow-line
regions (BLR and NLR). It has been suggested that the shifts of
certain lines may depend on orientation \citep[e.g.][]{Richards2002}
and that there is evidence for a flattened or disc-shaped BLR, with
the high-ionization lines being emitted closer to the central black
hole \citep[e.g.][]{Decarli2008}. See \citet{Marziani2008} for a
review of BLR models.  \citet{Popovic2004} suggest that there are two
components of the NLR, which could explain the blue shift of the
Lyman-$\upalpha$ narrow component relative to the [\ion{O}{iii}]
lines. In turn, it has been found that the [\ion{O}{iii}] line is
itself often blueshifted relative to H-$\upbeta$ \citep{Zamanov2002}
and the [\ion{O}{ii}], [\ion{N}{ii}], and [\ion{S}{ii}] lines
\citep{Boroson2005}.

Regardless of the physical origins of these shifts, we will have
understand their effects on the flux extrapolation in order to avoid
biasing the IGM measurements from high-redshift spectra. As we discuss
in \S \ref{halfFits} below, these shifts do indeed make it difficult
to accurately model the intrinsic spectra of high-redshift quasars. In
\S \ref{medianShiftSec} we attempt to correct for the average shifts,
but this relies on our knowledge of the intrinsic spectra of
low-redshift quasars. Without a clear understanding of the physical
origins of these shifts, any attempt to apply a similar correction to
the observed population of high-redshift quasars would be highly
uncertain. In particular, we worry that the shifts may be correlated
with redshift, or with quasar parameters such as luminosity, so that
the low-redshift sample is not necessarily representative of the
high-redshift population. However, in \S \ref{simulatedResSec} we show
that by careful choice of the wavelength range used in our analysis,
we reduce the bias induced by these shifts, and in \S
\ref{conclusionSec} we argue that the sign of the bias actually
strengthens current constraints on the IGM neutral fraction.

\subsection{Intrinsic asymmetry of the Lyman-$\upalpha$ emission}

\begin{figure}
 \includegraphics[width=3in]{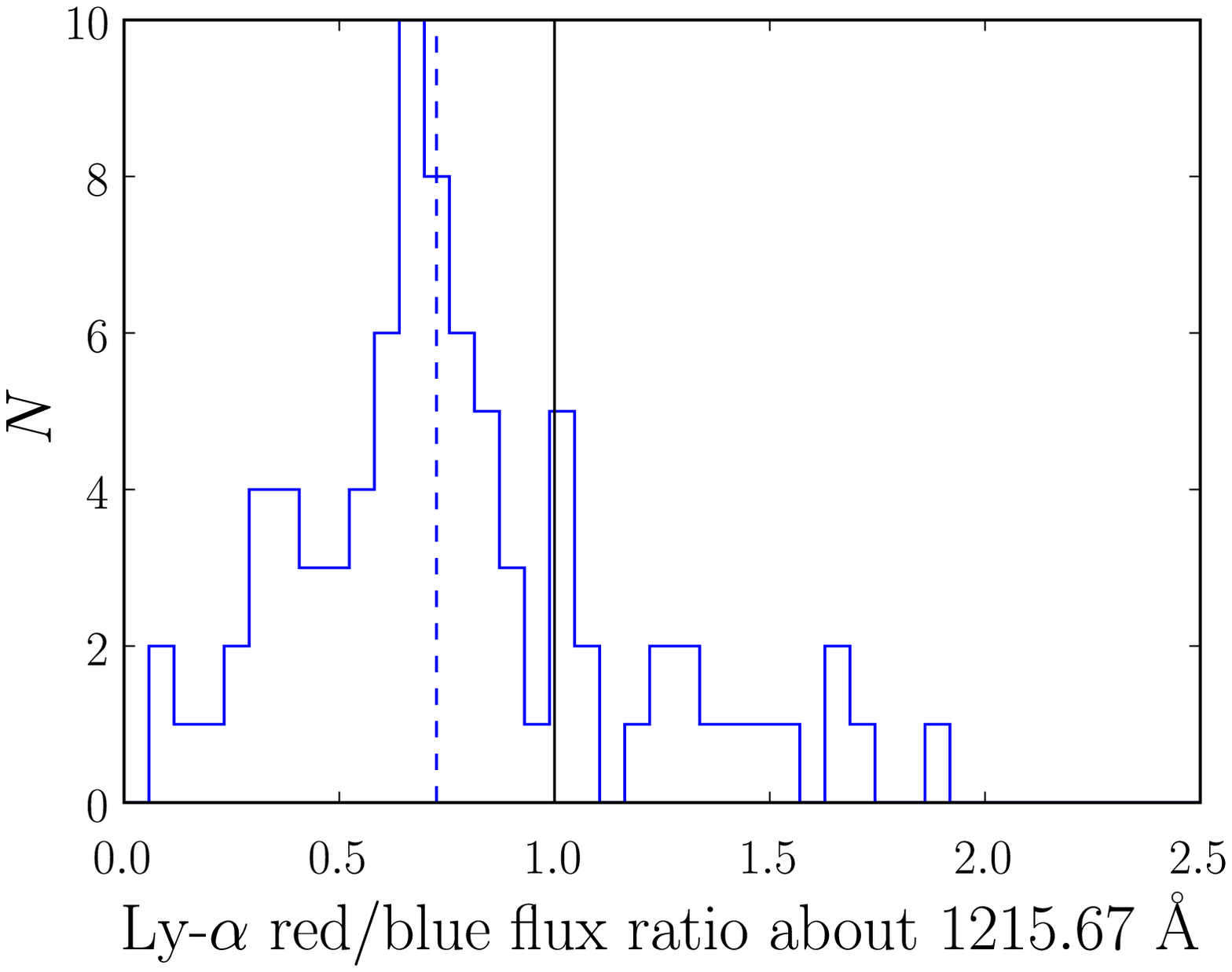}
 \includegraphics[width=3in]{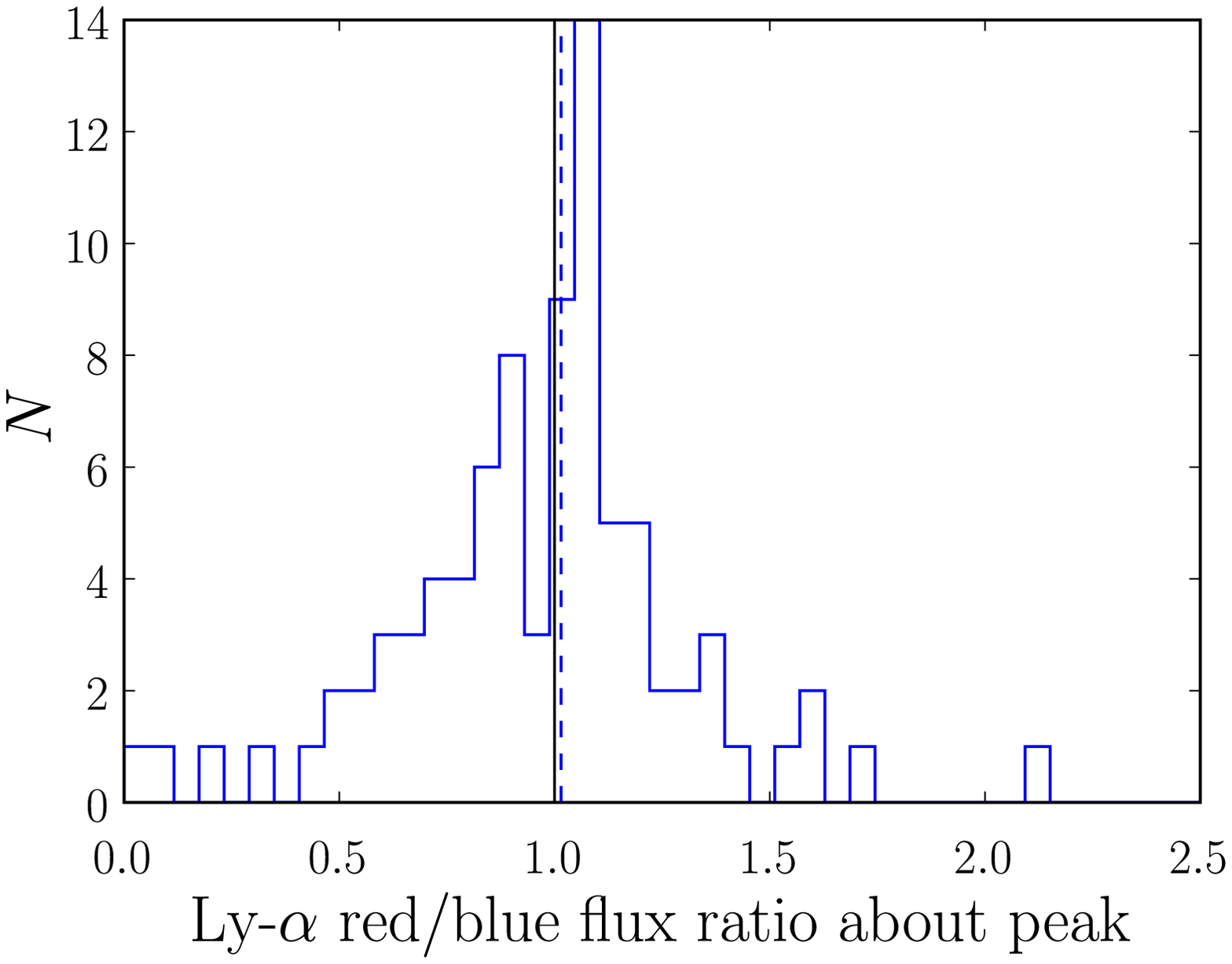}
 \caption{Distributions of ratio of red-side to blue-side
   Lyman-$\upalpha$ flux. In the top panel the flux is split at
   $(1+z)\;1215.67$~\AA\ in the observed frame. In the bottom panel the
   flux is split at the wavelength of the peak of the line. Median
   values (indicated by vertical dotted lines) are $0.73$ in the top
   panel and $1.02$ in the bottom. Three extreme values between 2.5
   and 7 fell outside of the range plotted in the top
   panel. \label{fluxRatioFig}}
\end{figure}

Another way to evaluate the asymmetry of the emission components (in
addition to the shifts of the best-fit-model emission components) is
to compare the amount of flux on the red and blue sides of the
line. Unfortunately, the \ion{N}{v} line makes it difficult to do this
in a model-independent way, so we rely on our line-profile fits to
distinguish between the Lyman-$\upalpha$ flux and the \ion{N}{v}
flux. In Figure \ref{fluxRatioFig} we plot the histogram of the total
flux on the red side of the Lyman-$\upalpha$ line to the total on the
blue side. The flux used is the sum of the narrow and broad
Lyman-$\upalpha$ components from the best-fit model (and therefore
excludes the continuum and \ion{N}{v} components). A value less than
one indicates an overall blueshift of the flux. In the top panel we
have divided the flux at the nominal Lyman-$\upalpha$ wavelength of
$\lambda_0 = 1215.67$~\AA. As we would expect from the biased
distribution of velocity shifts in Figure \ref{vshiftFig}, there is a
systematic excess of flux on the blue side of $1215.67$~\AA.

This invites the question of whether the Lyman-$\upalpha$ profiles are
symmetric about some other axis, or are they intrinsically asymmetric?
The peak of the line should define the axis of symmetry for a
symmetric line profile. The bottom panel of Figure \ref{fluxRatioFig}
shows the distribution of red-side to blue-side flux split at the
wavelength of the peak of the overall line profile. The peak was
calculated from the observed spectrum (not the model), by fitting a
quadratic to the 10 pixels surrounding the maximum observed flux value
and taking the peak wavelength of that function. As mentioned above,
this peak corresponds quite closely with the center of the narrow
component in all but a few cases (which generally show highly unusual
line morphology, often heavily blended with the \ion{N}{v} line or
having an absorption line near the peak). There is a wide spread in
this flux ratio distribution, but it is centered quite close to unity
(as measured by both the peak of the histogram and the median value,
indicated by the vertical dashed line). On average, the
Lyman-$\upalpha$ emission is symmetric about the peak of the line, but
in individual spectra the broad component can be displaced
considerably relative to the narrow component, producing an asymmetric
total line profile. In the next section we will explore the impact of
these shifts and asymmetries on our ability to accurately extrapolate
quasar spectra from the red side of the Lyman-$\upalpha$ line to the
blue side.

\section{Extrapolating to Obtain the Blue Flux from Fits to the Red Side Alone}
\label{halfFits}

In order to mimic fits to high-redshift quasar spectra (which are
subject to strong IGM absorption on the blue side of the
Lyman-$\upalpha$ line) we next perform fits using only the spectrum on
the red side of the line. Note that actual high-redshift spectra are
subject to some additional absorption on the red side of the line due
to the IGM damping wing. For simplicity, we do not include the
contribution of the damping wing in our continuum fits. In principle,
the continuum could always be self-consistently re-estimated for each
model hypothesis for the foreground IGM absorption. In practice, the
damping wing on the red side is nearly flat (constant with wavelength)
and will have a negligible impact on the continuum determination,
unless the universe is nearly neutral; furthermore, the continuum is
subdominant to the Lyman-$\upalpha$ emission line at the wavelengths
most important for our analysis. In order to make sure to avoid all of
the resonant absorption, including from any foreground gas falling
towards the quasar that would absorb light on the red side of the line
center \citep{BarkanaLoeb2003}, \citetalias{MH2007} excluded not only
the blue side of the line profile, but also the peak of the observed
line from their line-profile fits. We therefore limit our red-side
fits to $\lambda \ge 1220$~\AA\ in the rest frame.

\begin{table*}
\caption{Best-fit spectral parameters from fits to the red side only of the
Lyman-$\upalpha$ line.}\label{halfparamTable}

\medskip
\hrule
\bigskip
Complete table available in the electronic version.
\medskip
\hrule
\bigskip

The first two columns give the NED name of the target, plus the HST
instrument, grating, and aperture used to take the
spectrum. $\sigma_\mathrm{L\upalpha0}$ is the width of the narrow
Lyman-$\upalpha$ component (corrected for redshift using the NED value
of $z$, see Table 1). $a_\mathrm{L\upalpha0}$ is the amplitude of this
component, normalized by the continuum flux
$a_\mathrm{p}$. $v_\mathrm{L\upalpha0}$ is the shift of the component
center from the nominal wavelength calculated with the NED
redshift. The next six columns give the same quantities for the broad
Lyman-$\upalpha$ component ($\mathrm{L\upalpha1}$), and the \ion{N}{v}
component. $p$ is the continuum power law index, and $a_\mathrm{p}$ is
the continuum flux at the nominal Lyman-$\upalpha$ central wavelength,
in units of $10^{-15} \mathrm{erg~s^{-1}~cm^{-2}~\AA^{-1}}$. The last
columns give the reduced $\chi^2$ value $\chi^2/\nu$ and the number of
degrees of freedom $\nu$ for the continuum and line-profile fits.

\end{table*}

Without the leverage of flux data in the peak and blue wing of the
line there is not enough information to simultaneously constrain all
three Gaussian parameters for all three components in most cases. We
attempted to constrain the full set of line-profile parameters using
red-side-only fits, but found that the Gaussian components were often
poorly behaved and the resulting best-fit models were too poorly
constrained to be useful for extrapolating the flux on the blue side
of the line. \citetalias{MH2007} avoided this problem by fixing the
central wavelength of each component to the nominal wavelength of the
line using the systematic redshift determined from metal lines. As we
noted above, this is not a reliable assumption, but it is the most
reasonable conservative approach, and yielded better matches to the
blue-side spectral shapes than fits with free line centers. Similarly,
we found that the extrapolation of the continuum to the blue side of
the line was more reliable if we fixed the power law index to $-1.3$
in the red-side-only fits (and also in any full-profile fits to
spectra lacking coverage at $\lambda_0 < 1165$~\AA). Figure
\ref{specFitFig} (bottom plot) illustrates the type of error
introduced by extrapolating the flux using fixed line centers. In this
case the observed line profile is shifted blueward by $\sim
600~\mathrm{km/s}$. The amplitudes of the Gaussians in the red-side
fit with fixed centers are therefore depressed, and the flux is
underestimated on the blue side of the line. Table
\ref{halfparamTable} lists the best-fit model parameters for all of
the red-side-only fits (see description of Table
\ref{fullparamTable}).

\begin{figure}
  \includegraphics[width=3in]{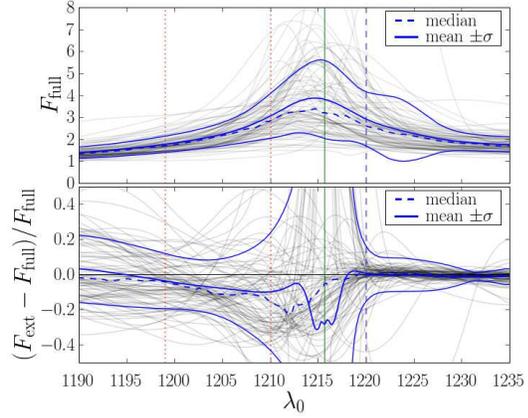}
  \caption{Fractional flux excess in the red-side-only fits versus
    wavelength. The excess is calculated by comparing the
    red-side-only best-fit model ($F_\mathrm{ext}$) to the
    full-profile fit ($F_\mathrm{full}$). The top panel shows the
    models fit to the full profiles of all 87 spectra (light grey) and
    the median (dashed) and mean and standard deviation (solid) of the
    set of model spectra. The bottom panel shows the fractional
    difference between the models fit to the red side only and the
    models fit to the full profile for all 87 spectra (light grey),
    and the median (dashed) and mean and standard deviation (solid) of
    the fractional flux difference. The vertical solid line (green)
    indicates the nominal central wavelength of the Lyman-$\upalpha$
    line. The vertical dashed line (blue) at $1220$~\AA\ indicates the
    blue edge of the red-side-only line profile fit region. The
    vertical dotted lines at $1199$~\AA\ and $1210$~\AA\ (red)
    demarcate the typical analysis region for the high-redshift IGM
    measurements. Note that within this wavelength range the
    extrapolation is more accurate at shorter
    wavelengths.}\label{fluxvarFig}
\end{figure}

\begin{figure}
  \includegraphics[width=3in]{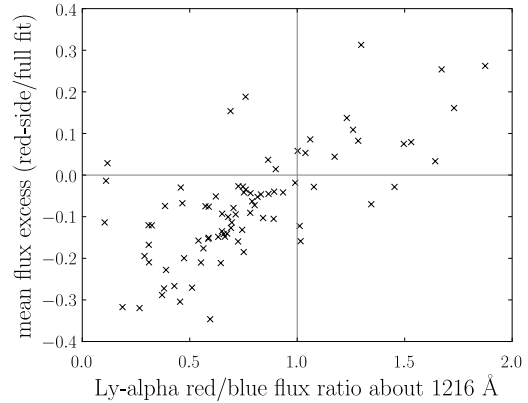}
  \caption{Mean fractional flux excess versus red/blue flux ratio of
    the Lyman-$\upalpha$ emission about $1215.67$~\AA. The fractional
    flux excess (see Fig. \ref{fluxvarFig}) is averaged over the
    region between $1199$~\AA\ and $1210$~\AA . The distribution of
    flux ratios is shown in the top panel of Figure
    \ref{fluxRatioFig}.}\label{fluxvarVsRatioFig}
\end{figure}

Figure \ref{fluxvarFig} shows the differences between full-profile and
red-side-only spectral fits for all 87 spectra. The top panel shows
the models fit to the full profiles of each spectrum together with the
median, mean, and standard deviation of the set of model spectra. The
vertical solid line indicates the nominal central wavelength of the
Lyman-$\upalpha$ line. The mean and median spectra clearly peak a few
\AA\ blueward of the nominal rest wavelength. The bottom panel shows
the fractional difference between the models fit to the red side only
and the models fit to the full profile for each spectrum, together
with the median, mean, and standard deviation of the fractional flux
difference. The vertical dashed line at $1220$~\AA\ indicates the blue
edge of the red-side-only line profile fit region. The agreement
between the fits is fairly good on the red side of this limit, as we
would expect, since both fits are using the same data here. At shorter
wavelengths not only are the differences larger, but there is a
bias. \textit{The spectra extrapolated from a model fit to the red
  side only of the Lyman-$\upalpha$ line tend to systematically
  underestimate the true spectra}.

Both the bias and spread of the flux extrapolation errors vary widely
with wavelength blueward of the $1220$~\AA\ lower limit on the
red-side-only spectral fits. The core of the Lyman-$\upalpha$ line is
quite poorly predicted from the red-side fits. This is partially
because the $1220$~\AA\ lower limit excludes the region where the
narrow component dominates, so this component is largely unconstrained
in these fits, which is why there is such a large variance in the
extrapolation error near line center. The main reason the bias is
largest close to the line center (but slightly blueward) is the fact
that the line centers are fixed in the red-side fits, while the real
lines (and thus the full-profile best-fit models) have emission
components that are often shifted by many \AA\ (as we saw in
Fig. \ref{vshiftFig}), often to shorter wavelengths. The fixed-center
Gaussians in the red-side models can match the red flank of the
emission-line fairly well, but once the line starts to peak and turn
over somewhere other than $1216$~\AA, the quality of the fit has to
decline. Specifically, if the real peak of the line is at $\lambda_0 <
1216$~\AA, then a Gaussian profile fit to the red side of the line
with a center fixed at $1216$~\AA\ will tend to underestimate the flux
on the blue side of the peak (as in Fig. \ref{specFitFig}). This
explains why the bias (as measured by the median error spectrum) is
close to zero between $1220$ and $1216$~\AA, but is largest just
blueward of $1216$~\AA\ where most of the observed lines peak. The
effect is exaggerated by the fact the line profile is most dominant
over the continuum at its peak. The bias is smaller farther to the red
where the continuum becomes more and more dominant, indicating that
the fixed powerlaw index of $-1.3$ is a reasonable \textit{mean} value
for this sample, though individual spectra still under- and
over-estimate the flux by large fractions. The slight positive bias at
$\lambda_0 < 1195$~\AA\ is due to the broad \ion{C}{iii}\* emission
line at $1175.7$~\AA.

The systematic underestimate of the flux by the red-side-only fits
will have important consequences for our ability to measure the IGM
neutral fraction. The vertical dotted lines at $1199$~\AA\ and
$1210$~\AA\ (red) indicate the typical analysis region used for the
high-redshift IGM measurements (see \S
\ref{analysisSubSec}).\footnote{We will be careful throughout this
  paper to distinguish the optical-depth fits used to measure IGM
  properties from high-redshift spectra from the line-profile fits
  used to study or extrapolate the intrinsic spectrum of a quasar.}
This IGM analysis range avoids the core of the Lyman-$\upalpha$ line,
where the extrapolation errors are largest but, nonetheless, on
average the red-side-only fits underestimate the flux over the entire
analysis region. Figure \ref{fluxvarVsRatioFig} shows the relationship
between the mean fractional flux excess in the analysis region and the
red/blue flux ratio of the Lyman-$\upalpha$ emission about
$1215.67$~\AA . As we expected, red-side-only fits to spectra with
blue shifted Lyman-$\upalpha$ flux tend to underestimate the blue-side
flux, and vice-versa. In \S \ref{simulatedResSec} we explore the
possibility of restricting the IGM analysis to $< 1205$~\AA\ to reduce
the bias, but this obviously also reduces the number of pixels
available to analyze. In \S \ref{medianShiftSec} we also attempt to
improve on our naive choice of $1216$~\AA\ for the Lyman-$\upalpha$
line centers with a crude correction for the mean blueshift of the
line components, though we caution again that this requires \textit{a
  priori} knowledge of (or inferences about) the distribution of line
shifts in the high-redshift quasar population.

Our primary concern in this paper is to understand how errors in flux
extrapolation like those described above (particularly those caused by
the asymmetries in the line profiles) propagate to errors in the final
determination of the IGM neutral fraction. In the next section we
describe how our line-profile fits are used to simulate high-redshift
damping wing measurements in order to evaluate this impact.

\section{Simulated high-$z$ damping wing measurements}\label{simulatedSec}

\subsection{Generating simulated high-$z$ quasar spectra}

In order to understand how flux extrapolation errors will affect our
ability to measure the IGM neutral fraction using the IGM damping
wing, we must simulate high-redshift quasar spectra. We begin with the
full-profile spectral model fits described in \S \ref{fullFits}. Each
best-fit model is used as the intrinsic spectrum of a hypothetical
high-$z$ object. We use the line-profile models instead of the actual
observed spectra of the low-$z$ quasars so that we can match the
expected noise and instrumental resolution of a high-$z$ spectrum. In
order to make sure that our flux models capture all of the features
relevant to the IGM neutral fraction determination, we also carried
out an analysis using the observed spectra (with flagged absorption
features, but not emission features, replaced by the model flux
values) as the intrinsic spectra of the simulated high-$z$ quasars. We
found that the results did not differ significantly from our analysis
using the model line profiles as the intrinsic spectra.

We then simulate absorption by neutral hydrogen in the IGM (and
instrumental effects) to generate mock absorption spectra on which we
can test our techniques for recovering the IGM parameters. We model
the density field along lines of sight through the high-$z$ IGM using
the probability distribution function (PDF) of \citet*{MHR2000}:
\begin{equation}\label{pdfEq}
  P_V(\Delta) d\Delta = A \Delta^{-\beta} ~ %
  \exp{\left[%
      \frac{-(\Delta^{-2/3} - C)^2}%
           {2~(2 \delta_0/3)^2}%
           \right]} d\Delta,
\end{equation}
where $\Delta \equiv (\rho / \rho_0)$ is the density in the IGM
normalized by the mean density, and we use the $z=6$ values $A =
0.864$, $\delta_0 = 1.09$, $\beta = 2.5$, and $C = 0.880$ for all of
the constants. Our conclusions should be fairly insensitive to small
changes in these values, or even in the form of the PDF. The crucial
assumption we make is that the density PDF is fairly well known. For
actual measurements, uncertainties in the form of the PDF would be an
additional source of error that we do not address here.

To construct density profiles $\Delta(r)$ along the line of sight, we
generated independent random values of the density from Equation
\ref{pdfEq} for each patch of IGM. In order to roughly account for
photoionization-induced smoothing in the density-field
\citep{Gnedin2000}, we set a minimum co-moving size of
$1~\mathrm{Mpc}$ on IGM density patches. All pixels falling within
the same isodensity patch receive the same IGM density value. We do
this in order to avoid having an unrealistically high number of
independent data points in our simulated high-resolution spectra. In
our mock absorption spectra we have roughly 2 pixels per isodensity
patch, and roughly 20 patches are included in each damping-wing
analysis. The details of the IGM correlations should not matter a
great deal as long as they are short-range (a small fraction of the
size of the ionized region), especially since the instrumental FWHM we
use is comparable to the isodensity patch size. We have tested this
assumption by repeating the analysis with isodensity patches of
$0.25~\mathrm{Mpc}$ co-moving. The scatter is reduced slightly due to
the larger number of independent samples of the optical depth profile,
but our overall conclusions are unaffected.

Next we calculate the neutral hydrogen density assuming ionization
equilibrium. Three important parameters enter into the calculation at
this stage. The ionized region size is $R_\mathrm{HII}$. Outside of
this region, the ionizing flux is uniform, and is parametrized by
$x_\mathrm{IGM}$, the equilibrium neutral fraction at mean IGM
density. Inside the ionized region, there is additional flux from an
ionizing point source (the quasar), parametrized by $x_\mathrm{ref}$,
which is the equilibrium neutral fraction (including both background
and quasar flux) of mean-density gas at a reference point
$r_\mathrm{ref} = 35~\mathrm{Mpc}$ comoving away from the quasar.

Physically, the radius $R_\mathrm{HII}$ is set by the quasar's
ionizing photon luminosity and its age. The higher the rate of
ionizing photon production and the longer the quasar has been shining,
the larger its ionized region will be (at least until the
recombination rate inside the sphere balances the rate of photon
emission, which occurs only at times much longer than the expected
quasar lifetime \citealt{CH2000}). The neutral fraction at a given
point inside the ionization front $x_\mathrm{ref}$ scales with the
luminosity of the quasar.

We define a normalized ionization rate
\begin{equation}
\gamma \equiv \frac{\Gamma}{n_\mathrm{H} \upalpha_B},
\end{equation}
where $\Gamma$ is the number of ionizations per unit time per neutral
hydrogen atom, $n_\mathrm{H}$ is the total (neutral and ionized)
hydrogen density, and
$\upalpha_B=2.59\times10^{-13}~\mathrm{cm^3~s^{-1}}$ is the case B
recombination coefficient at $T = 10^4~\mathrm{K}$. The normalized
background ionization rate required to maintain an equilibrium neutral
fraction of $x_\mathrm{IGM}$ is then
\begin{equation}
\gamma_\mathrm{BG} = \frac{(1 - x_\mathrm{IGM})^2}{x_\mathrm{IGM}}.
\end{equation}
In order to maintain a neutral fraction $x_\mathrm{ref}$ at the
reference distance, the additional normalized ionization rate due to
the quasar flux must be
\begin{equation}
\gamma_\mathrm{ref} = \frac{(1 - x_\mathrm{ref})^2}{x_\mathrm{ref}} -
\gamma_\mathrm{BG}.
\end{equation}
Therefore the total ionization rate as a function of luminosity distance from
the quasar $r$ (inside the ionized region), is
\begin{equation}
\gamma(r) = \frac{\bar{n}_\mathrm{H}(r_\mathrm{ref})}{n_\mathrm{H}(r)} 
\left[ 
  \gamma_\mathrm{ref} \left(\frac{r_\mathrm{ref}}{r}\right)^2 + \gamma_\mathrm{BG}
  \right],
\end{equation}
where the ratio of the mean IGM hydrogen density at $r_\mathrm{ref}$
to the local hydrogen density at $r$ is given by
\begin{equation}
\frac{\bar{n}_\mathrm{H}(r_\mathrm{ref})}{n_\mathrm{H}(r)} = \Delta(r)^{-1}~\left[ \frac{1
    + z(r_\mathrm{ref})}{1 + z(r)}\right]^3.
\end{equation}
This factor simply compensates for the fact that $\gamma_\mathrm{ref}$
and $\gamma_\mathrm{BG}$ are defined with reference to the mean IGM
density at $r_\mathrm{ref}$. Outside the ionized region ($r >
R_\mathrm{HII}$), the quasar contribution is zero
($\gamma_\mathrm{ref} = 0$), so the ionization rate is simply
\begin{equation}
\gamma(r) = \frac{\bar{n}_\mathrm{H}(r_\mathrm{ref})}{n_\mathrm{H}(r)} \gamma_\mathrm{BG}.
\end{equation}
The equilibrium neutral fraction at distance $r$ from the quasar is
\begin{equation}
x(r) = \frac{ -b - \sqrt{b^2 - 4}}{2},
\end{equation}
with
\begin{equation}
b = -2 - \gamma(r).
\end{equation}

Finally, the optical depth due to resonant absorption is
\citep{Miralda-Escude1998}
\begin{equation}
  \tau_\mathrm{r}(\lambda) = \tau_0 ~ x(r) %
  \left[\frac{\Omega_b h_0 X}{0.03}\right] %
  \left[\frac{H_0 (1 + z)^{3/2}}{H(z)}\right]%
  \left[\frac{1 + z}{6}\right]^{3/2},
\end{equation}
where $\tau_0 = 2.1 \times 10^5$, $r$ and $z$ are the distance and
redshift corresponding to wavelength $\lambda$, $H(z)$ is the Hubble
constant at redshift $z$ \citep[see, e.g.][]{Hogg1999}, and $h_0 =
H_0/100~\mathrm{km~s^{-1}~Mpc^{-1}}$. We use $\Omega_b = 0.0462$,
$\Omega_c = 0.233$, $\Omega_\Lambda = 0.721$, $\Omega_m = \Omega_c +
\Omega_b$, $\Omega_k = 0$, $h_0 = 0.701$, and the hydrogen fraction $X
= 1- 0.240$ \citep{Komatsu2008}.

Inside the transmission window in the quasar spectrum (corresponding
to the ionized region), at wavelength $\lambda$ there is both resonant
absorption $\tau_\mathrm{r}$ by hydrogen at $z = \lambda/\lambda_0 -
1$, and absorption by the extended damping wing $\tau_\mathrm{d}$ of
hydrogen at lower redshifts in the Gunn-Peterson trough. We calculate
$\tau_\mathrm{d}$ using an analytical expression for the red wing of
the GP trough derived by \citet[][equations 11 and
  12]{Miralda-Escude1998}.

The total optical depth inside the transmission window is the sum of
the resonant contribution from the residual neutral hydrogen inside
the ionized region, and the damping wing absorption from the hydrogen
in the IGM,
\begin{equation}
\tau(\lambda) = \tau_\mathrm{r}(\lambda) + \tau_\mathrm{d}(\lambda),
\end{equation}
and the transmitted flux is
\begin{equation}\label{fluxEq}
F_\mathrm{obs}(\lambda) = F_\mathrm{i}(\lambda) e^{-\tau(\lambda)},
\end{equation}
where $F_\mathrm{i}(\lambda)$ is the intrinsic spectrum of the quasar.

Finally, we smooth the mock spectrum with an instrumental line spread
function and add noise. We use a spectral resolution of $R =
2000$. The spectrum is smoothed by Gaussian convolution with a kernel
of width $\sigma_\mathrm{LSF} = \lambda_0 (1+z) / (2.355~ R)$. The
pixel size is $\Delta\lambda = \lambda_0 (1+z) / (2.5~R)$. We
normalize the spectral models to have the same intrinsic continuum
flux $F_0$ at $\lambda_0$. Our noise model, chosen by hand to mimic a
typical spectrum of a high-$z$ quasar \citep{Fan2006a}, is a
combination of a shot-noise term and a constant background term,
giving flux variance (in normalized units) $\sigma_F^2(\lambda) =
e_F^2 F(\lambda) + e_\mathrm{BG}^2$ with $e_F = 10^{-2}$ and
$e_\mathrm{BG} = 10^{-3}$. This rough, order of magnitude estimate of
the noise level suffices, since, in practice, instrumental noise has
very little impact on the fits due to the much larger fluctuations in
the observed flux caused by density variations in the IGM.

\begin{figure}
    \includegraphics[width=3in]{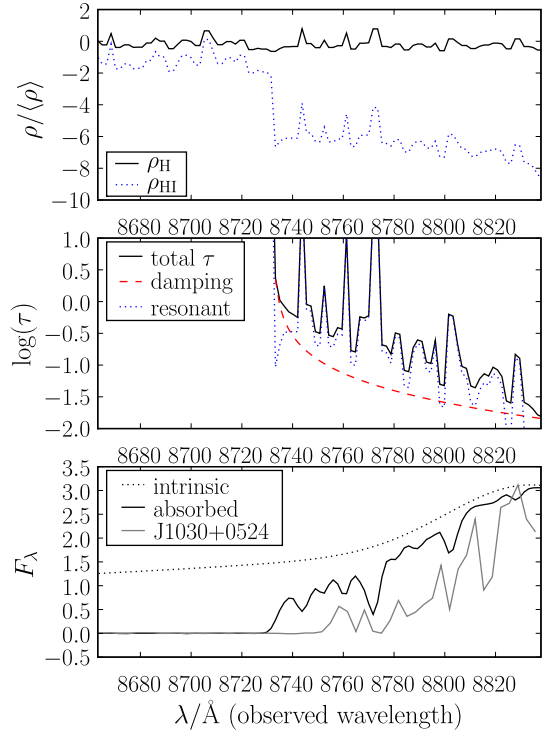}
    \caption{\label{densityFig} Generating a mock absorption spectrum:
      density (top), optical depth (middle), and flux (bottom)
      profiles along the line of sight to a hypothetical $z = 6.28$
      quasar. The top panel shows the total ($\rho_\mathrm{H}$, black)
      and neutral ($\rho_\mathrm{HI}$, blue dotted) hydrogen density
      normalized to the mean density of the IGM. The model parameters
      are $R_\mathrm{HII} = 40.5~\mathrm{Mpc}$, $x_\mathrm{ref} =
      10^{-5.5}$, and $x_\mathrm{IGM} = 0.1$. The middle panel shows
      the damping wing and resonant optical depths, and their sum. The
      extreme optical depths in the Gunn-Peterson trough (at $\lambda
      \la 8730$~\AA) are above the range of this plot. The bottom
      panel shows a model intrinsic spectrum (dotted) and mock
      absorption spectrum (solid black). A Keck spectrum of the
      $z=6.28$ quasar J1030+054 (grey) is also shown
      \citep{Becker2001}. Note that the model was not fit to this
      spectrum, it is shown merely for comparison.}
\end{figure}

Figure \ref{densityFig} shows the density, optical depth, and flux
profiles for a model high-redshift quasar. The effects of both the
uniform ionizing background and the quasar flux are evident in the
deficit in the neutral hydrogen density relative to the total hydrogen
density. The optical depth is much lower inside the ionized region,
allowing the quasar flux to reach us. With these parameters, the
resonant absorption dominates the opacity in high-density regions,
while the damping wing dominates in low-density regions, setting an
effective optical depth floor. The plot of the spectrum (in the bottom
panel) demonstrates that the transmission window is superimposed on
the blue wing of the quasar's intrinsic Lyman-$\upalpha$ line. The
flux is entirely absorbed in the Gunn-Peterson trough (at $\lambda \la
8730$~\AA), but a large fraction is transmitted in the wavelength
range corresponding to ionized region. The Keck spectrum of the
$z=6.28$ quasar J1030+054 is qualitatively quite similar, differing
mostly in that the transmission window is slightly smaller and
instrumental broadening has smoothed the flux variations somewhat.

\subsection{Analysis of the mock spectra}\label{analysisSubSec}

\begin{figure}
  \includegraphics[width=3in]{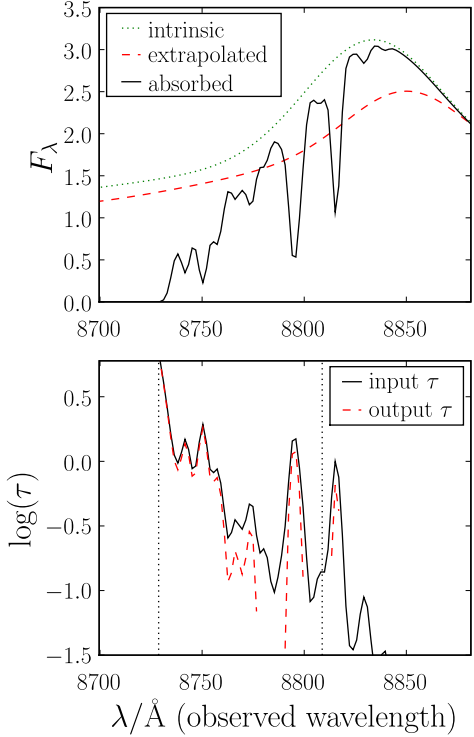}
  \caption{Error in the mock ``observed'' optical depth profile: Flux
    (top) and optical depth (bottom) on the blue side of the
    Lyman-$\upalpha$ line. The top panel shows the intrinsic
    (unabsorbed) flux, flux extrapolated from the red-side fit, and
    mock absorption spectrum for a model quasar at $z=6.28$ (based on
    the observed spectrum of 3C~273). The bottom panel shows the input
    model optical depth (used to generate the spectrum), and the
    output optical depth inferred by comparing the mock absorption
    spectrum and the extrapolated flux. The mismatch between these
    lines is caused by the mismatch between the intrinsic spectrum of
    the quasar, and the spectrum extrapolated from a fit to the red
    side of the Lyman-$\upalpha$ line. The vertical dotted lines (at
    $1199$~\AA\ and $1210$~\AA\ in the quasar rest frame) indicate the
    typical analysis region for the high-redshift IGM
    measurements.\label{depthFig}}
\end{figure}

In order to test the impact of intrinsic Lyman-$\upalpha$ line shape
variations on the recovery of IGM and \ion{H}{ii} region parameters
from a high-$z$ quasar spectrum, we implement a version of the method
used by \citetalias{MH2007}. The first step is to extrapolate the flux
by fitting a spectral model to the red side of the Lyman-$\upalpha$
line, as described in \S \ref{fitSec}. We then calculate the optical
depth by comparing the extrapolated spectrum to the mock absorption
spectrum and solving equation \ref{fluxEq} for the optical depth,
substituting the extrapolated flux $F_\mathrm{fit}$ for the intrinsic
flux and the observed flux $F_\mathrm{obs}$ for the transmitted flux:
\begin{equation}\label{tauObsEq}
\tau_\mathrm{obs} = \ln\left(\frac{F_\mathrm{fit}}{F_\mathrm{obs}}\right).
\end{equation}
An overestimate of the intrinsic flux results in an overestimate of
the optical depth, and vice-versa. Figure \ref{depthFig} illustrates
this effect. The top panel shows the intrinsic spectrum of a mock
high-redshift quasar (based on the observed spectrum of 3C~273), the
mock spectrum after absorption in the IGM, and the flux extrapolated
from a fit to only the red side of the Lyman-$\upalpha$ line. It can
be seen that the flux extrapolated from red-side-only fit in this case
(as is typical) underestimates the intrinsic spectrum of the model
quasar. When the extrapolated flux is used to calculate the optical
depth profile (shown in the bottom panel), the error in the flux
extrapolation results in errors in the inferred optical depth
profile. Within the analysis region typically used for high-redshift
IGM measurments ($1199$--$1210$~\AA\ in the rest frame), there is a
systematic underestimate of the optical depth which is largest at
longer wavelengths, where the extrapolation error is largest. Note
that there are regions of the spectrum where the absorbed flux exceeds
the extrapolated flux by significant fraction. This is outside of the
line-profile fit region as we have defined it here for high-$z$ fits
($\lambda_0 > 1220$~\AA), but in principle observed absorbed flux
values on the blue side of the line could be included in the fits as
lower limits on the intrinsic flux, which would decrease the
underestimate of the flux somewhat.

Once we have an optical depth profile, we need to fit a model to it to
recover our parameters of interest. Because the density fluctuations
in the IGM cause large flux variations, this is not a trivial
task. \citetalias{MH2007} solved the problem by grouping the flux
values into three wavelength bins, then using the Kolmogorov-Smirnov
(K-S) test to compare the distribution of flux values in each bin with
a distribution derived from simulations. The product of the three K-S
probabilities $p$ then served as the relative likelihood estimate for
that combination of parameters. They then compared observed quasars to
a grid of models using this technique, selecting the model with the
maximum $p$-value as the best fit. Optical depth values above about
$6$ are not measurable from a real spectrum because the flux drops
below the detection limit. We therefore set any $\tau > 6$ to
$\tau_\mathrm{max} = 6$. Note that the K-S test is not strictly valid
with censored data, and assumes all values are independent, neither of
which holds in this case. Therefore care must be taken in interpreting
the $p$-values. We rely on Monte-Carlo simulations to avoid depending
on the $p$-values to determine confidence intervals.

We will distinguish between our grid of ``model'' optical depth
profiles, used to make the canonical distributions for the K-S test,
and our set of ``mock'' profiles, which we are analyzing as if they
were observed high-$z$ quasars, even though they are all generated
using the same techniques outlined above. The only difference is that
the model optical depth profiles are generated without reference to a
spectrum, while the mock profiles use the extrapolated spectra, and
thus incorporate any errors in the flux extrapolation.

At each point in our parameter-space grid we generate 100 model
profiles, and we use each real low-$z$ quasar spectrum to generate
$200$ mock optical depth profile observations using a single set of
input parameters, making a total of $86 \times 200 = 17200$ mock
profiles. We then run the optical depth analysis on each mock
spectrum, and track the distribution of the recovered best-fit
parameters $(R_\mathrm{HII}, x_\mathrm{ref}, x_\mathrm{IGM})$.

By default, the maximum wavelength of the analysis range is
$\lambda_\mathrm{max} = 1210$~\AA\ in the rest
frame. \citetalias{MH2007} excluded pixels closer to the
Lyman-$\upalpha$ line center to avoid the biased environment close to
the quasar. This has the added advantage of avoiding the most
problematic area for flux extrapolation --- the
narrow-component-dominated core of the line (since the narrow
component is less well constrained in the line profile fits and is
affected more by shifts in the line center). The minimum wavelength is
chosen by finding the bluest pixel with $\tau \le 6$, then extending
the overall wavelength range blueward by $5\%$. The inclusion of these
extra ``dark'' pixels is important for the statistical comparison with
mock spectra, since density variations change the exact wavelength at
which the optical depth exceeds $\tau = 6$ for different spectra with
the same value of $R_\mathrm{HII}$. With $R_\mathrm{HII} =
40.5~\mathrm{Mpc}$, the typical range is roughly $1199$--$1210$~\AA,
divided evenly into three bins of about $15$ pixels each.

\section{The Impact on Parameters Estimated from High-$z$
  Spectra}\label{simulatedResSec}

\subsection{Parameter Recovery with Known Intrinsic Spectra}

\begin{figure}
  \includegraphics[width=3in]{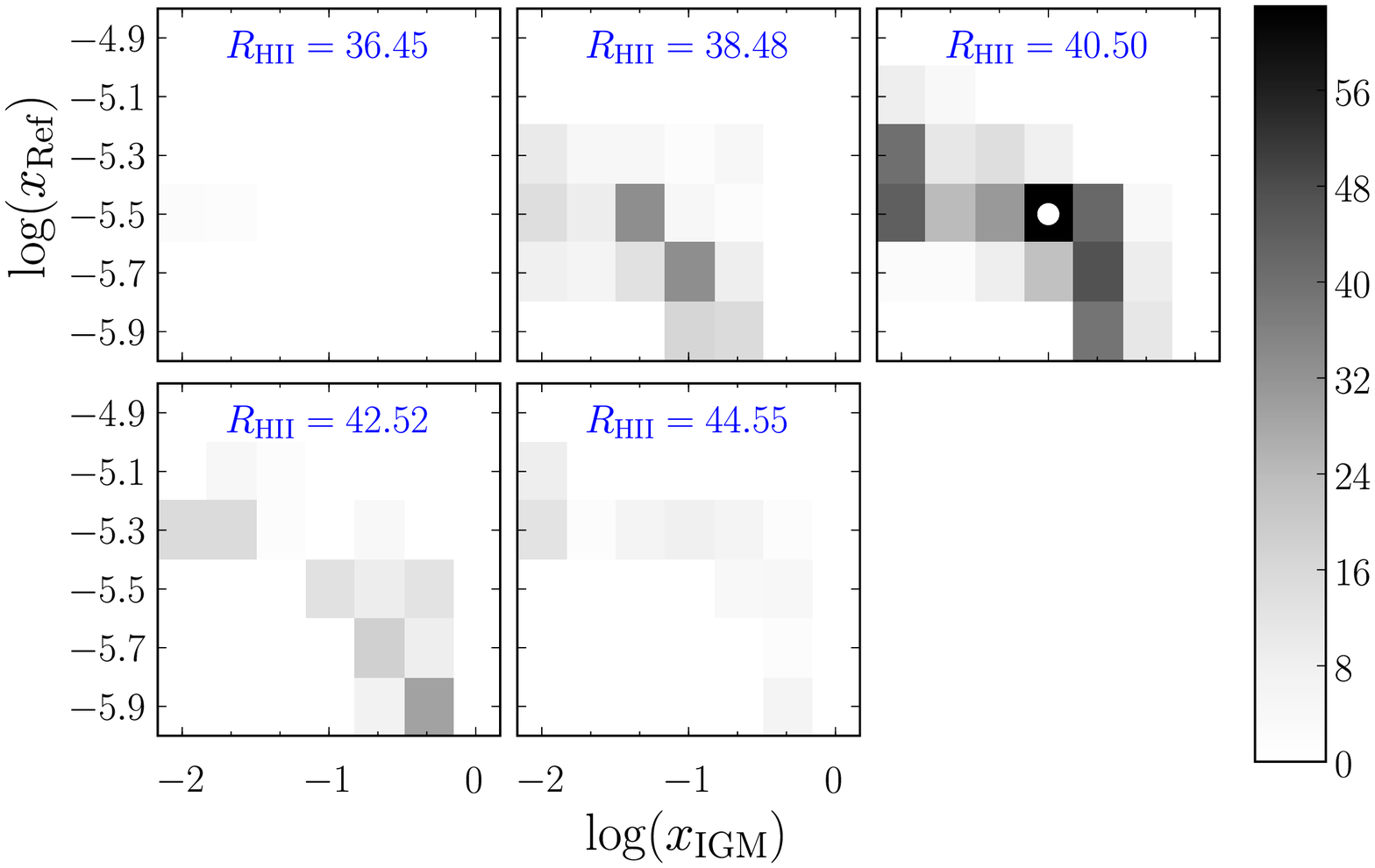}
  \includegraphics[width=3in]{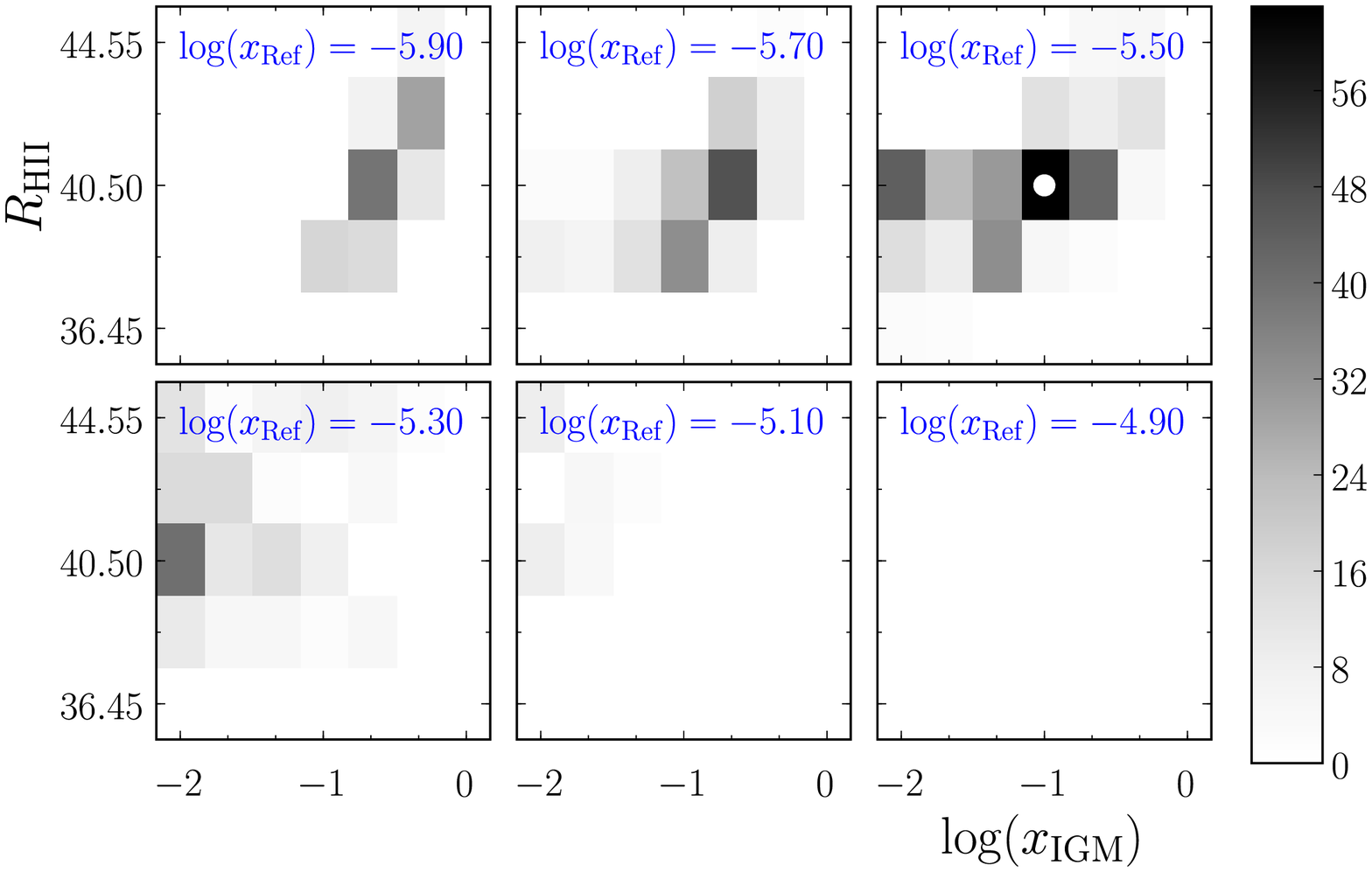}
  \includegraphics[width=3in]{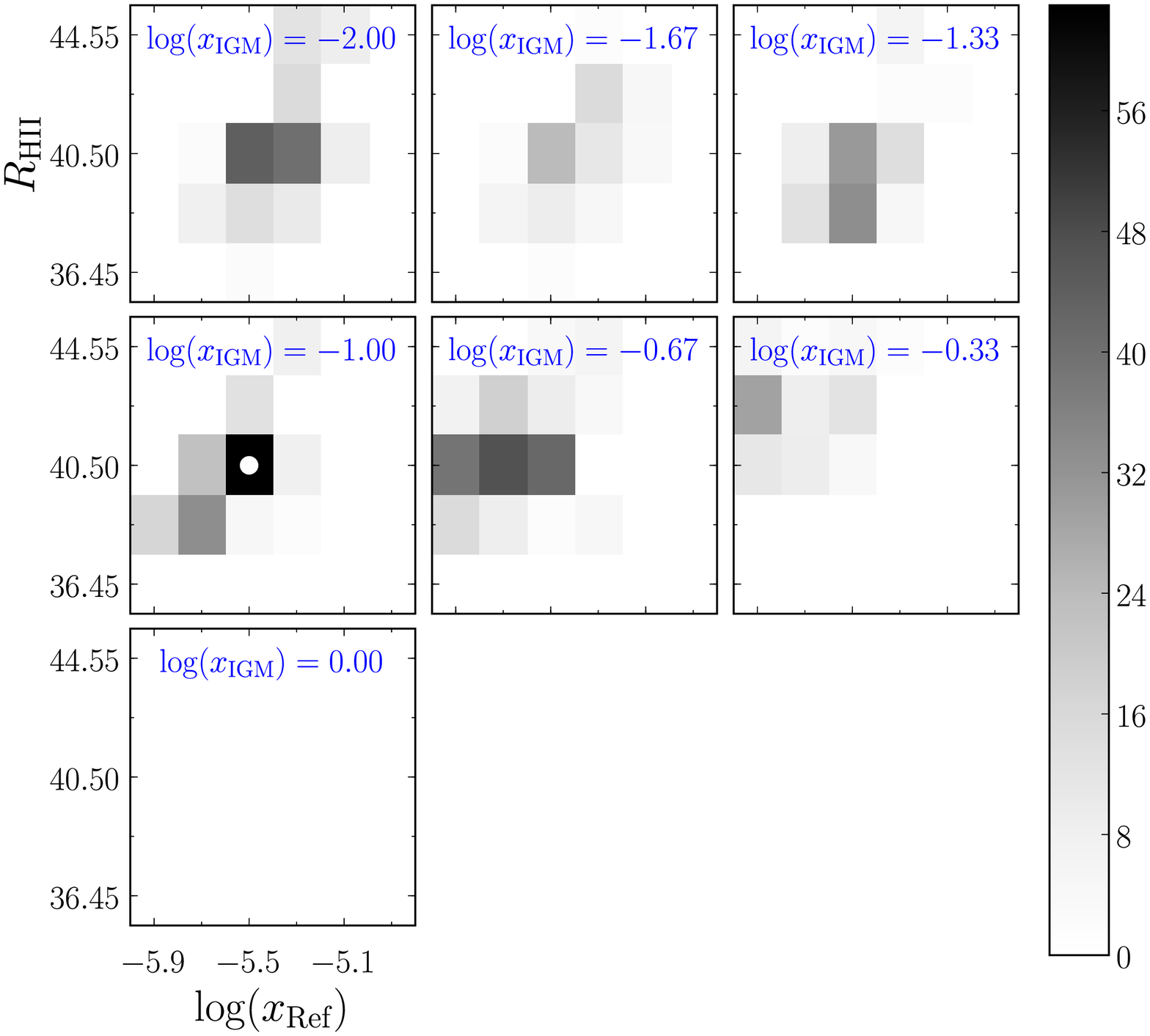}
  \caption{Parameter recovery map using perfectly known intrinsic
    spectra. Each panel is a slice through parameter space. The
    shading of each square indicates the number of best-fit values
    that fell within that cube in parameter space (darker indicates
    larger numbers). The input parameters
    ($R_\mathrm{HII}=40.5\ \mathrm{Mpc}$, $x_\mathrm{ref} =
    10^{-5.5}$, $x_\mathrm{IGM} = 0.1$) are indicated with a white
    point. The top group of graphs show slices of constant \ion{H}{ii}
    region radius. The middle group shows slices of constant internal
    reference neutral fraction. The bottom group shows slices of
    constant IGM neutral fraction. Note that the distribution peaks at
    the correct values, however there is a large scatter induced by
    density fluctuations along the line of
    sight.\label{paramSpacePerfectFig}}
\end{figure}

In order to test recovery of IGM parameters \textit{without}
extrapolation errors, as a sanity check we first calculate a set of
mock optical depth profiles using the ``correct'' intrinsic model
spectra in Equation \ref{tauObsEq} instead of the extrapolated spectra
$F_\mathrm{fit}$. We then analyze the profiles as described above. We
used only 4 different intrinsic spectral models (chosen at random from
our set of full-profile fits) for this purpose, since the shape of the
underlying spectrum has little impact on the parameter recovery if it
is perfectly known. Figure \ref{paramSpacePerfectFig} shows the
distribution of the best-fit values from the 800 mock optical-depth
fits in our 3-dimensional parameter space (200 random IGM density
profiles times 4 intrinsic spectra). The input parameters are
$R_\mathrm{HII} = 40.5~\mathrm{Mpc}$; $x_\mathrm{ref} = 10^{-5.5}$;
$x_\mathrm{IGM} = 0.1$. The shading of each square indicates the
number of best-fit values that fell within that cube in parameter
space (darker shading indicates larger numbers). The top group of
graphs show slices through our parameter space in planes of constant
$R_\mathrm{HII}$. The middle group shows slices of constant
$x_\mathrm{ref}$. The bottom group shows slices of constant
$x_\mathrm{IGM}$. The regions of darker shading trace out degeneracies
between all of the parameters, which are easily understood. Increasing
either $x_\mathrm{IGM}$ or $x_\mathrm{ref}$ increases the optical
depth, so these parameters are anti-correlated (top group of
graphs). When the best-fit value of the IGM neutral fraction is
overestimated, the internal neutral fraction tends to be
underestimated. On the other hand, increasing $R_\mathrm{HII}$ moves
the damping wing away from a given wavelength coordinate and decreases
the optical depth, so the ionized region radius is positively
correlated with the other two parameters (middle and bottom groups of
graphs).

\begin{figure}
  \includegraphics[width=3in]{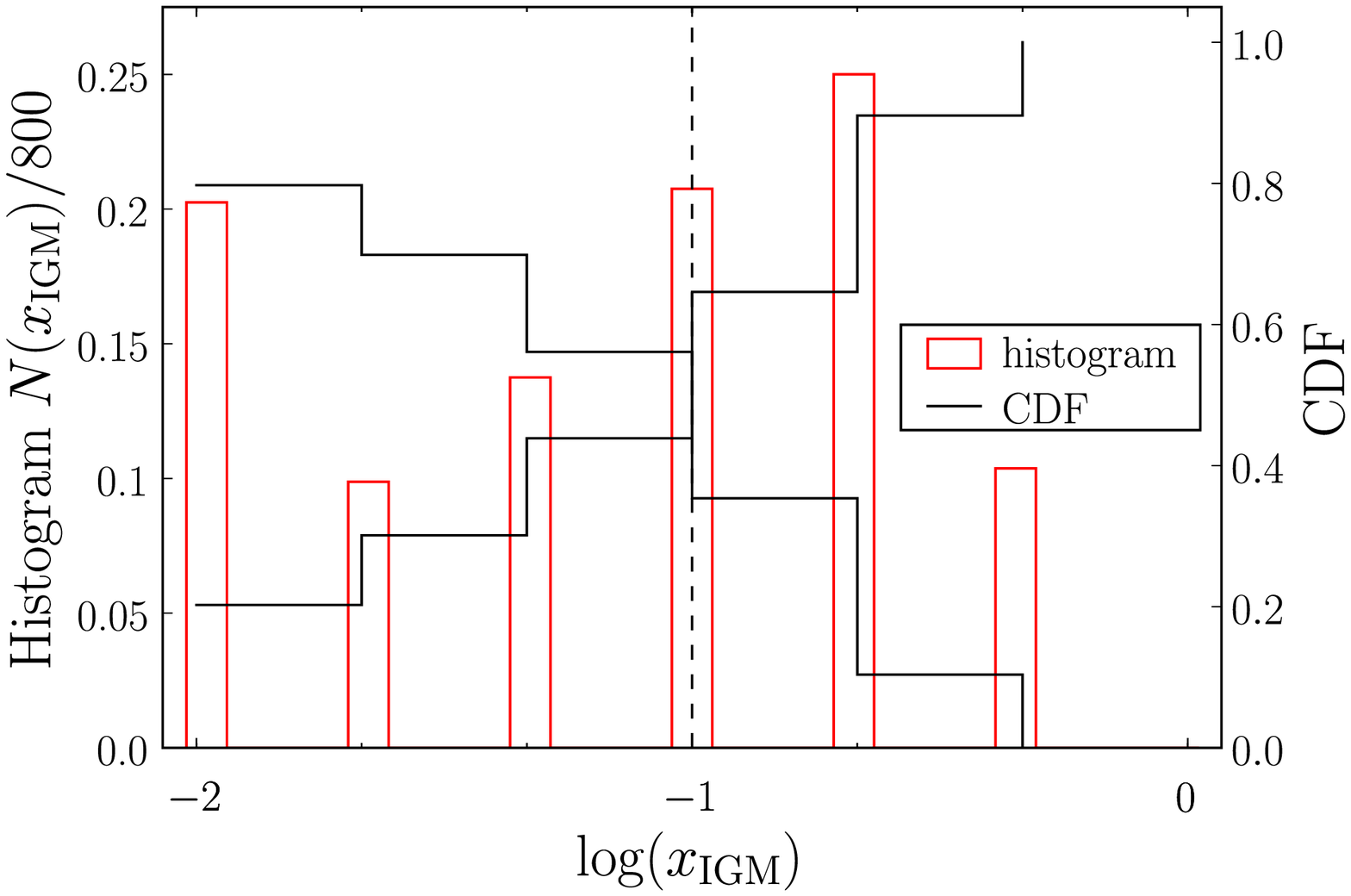}
  \includegraphics[width=3in]{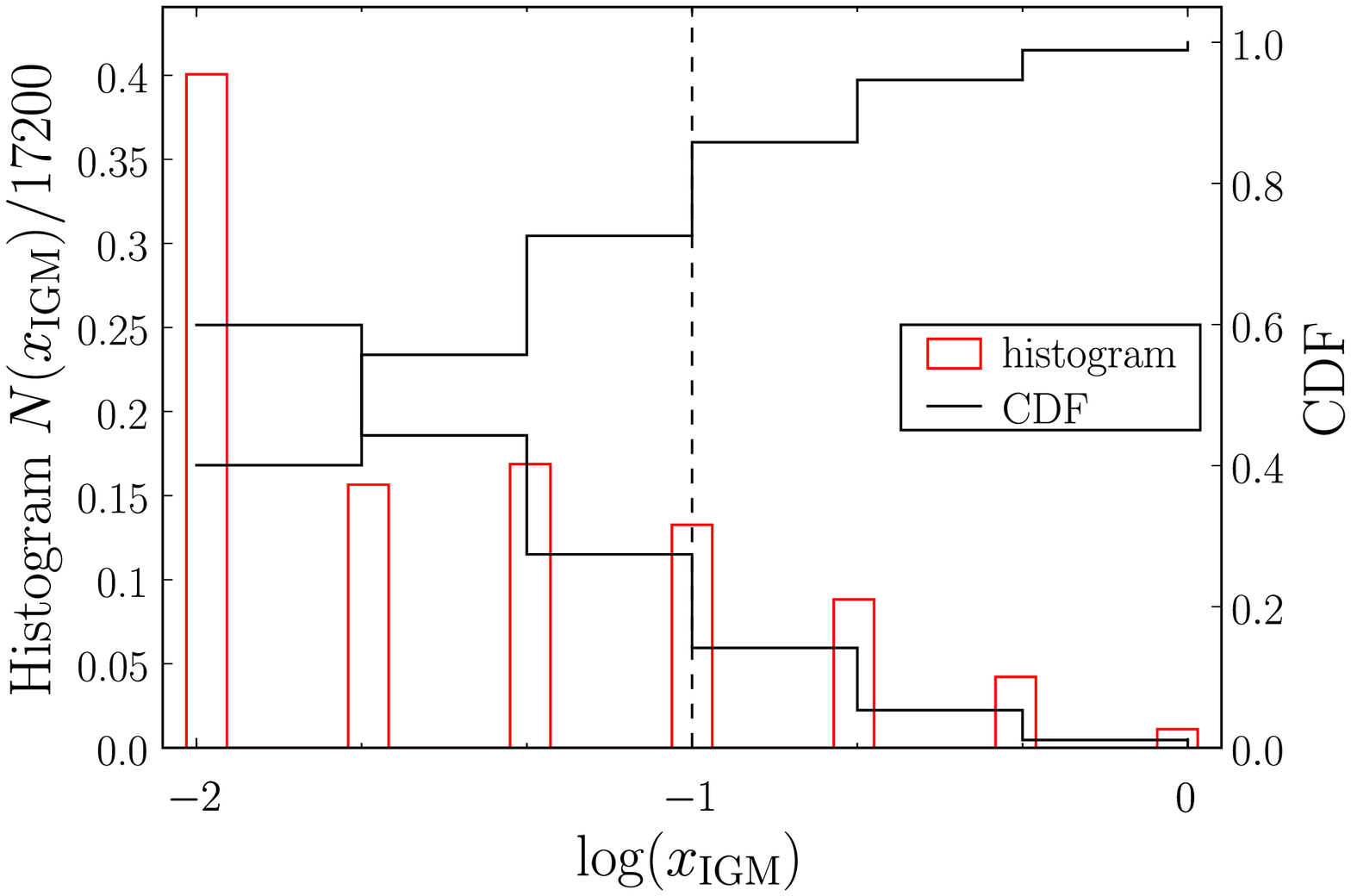}
  \caption{Marginalized distribution of estimated $x_\mathrm{IGM}$
    values with perfectly known (top) and extrapolated (bottom)
    intrinsic spectra. The histogram (axis scale on the left) shows
    the marginalized distribution of recovered $x_\mathrm{IGM}$
    values. Black lines show the cumulative distribution function
    (CDF), i.e. the fraction of recovered values at $x <
    x_\mathrm{IGM}$ (increasing curve, axis scale on the right) and
    its complement (decreasing curve). Input parameters are
    $R_\mathrm{HII}=40.5~\mathrm{Mpc}$, $x_\mathrm{ref} = 10^{-5.5}$,
    $x_\mathrm{IGM} = 0.1$. The vertical dashed line indicates the
    input value. The median value occurs where the CDF and its
    complement cross.\label{xIGMFig}}
\end{figure}

The density fluctuations along the line of sight induce a wide scatter
in the recovered values, which follows the degeneracy contours we
described above. Even so, the distribution of best-fit parameters in
Figure \ref{paramSpacePerfectFig} is peaked at exactly the coordinates
of the input parameters. The top panel of Figure \ref{xIGMFig} shows
the marginalized distribution of recovered $x_\mathrm{IGM}$
values. Black lines show the cumulative distribution function (CDF)
and its complement. Where they cross (at $50\%$) is the median
value. While the marginalized distribution peaks slightly above the
input value, the median recovers the input value (indicated by the
vertical dashed line) exactly. The secondary peak at
$\log(x_\mathrm{IGM}) = -2.0$ represents the integral of the long
negative tail of the distribution which is truncated by our finite
logarithmic grid (which we must keep small for computational speed).

Because of our coarse parameter-space grid it is difficult to quote
conventional confidence limits, but we can see from Figure
\ref{xIGMFig} that $80\%$ of the marginalized neutral fraction
distribution is in the interval $\log(x_\mathrm{IGM}) = -1.0 \pm 0.6$,
$90\%$ of the distribution is at $x_\mathrm{IGM} \leq 0.22$, and
$100\%$ at $x_\mathrm{IGM} \leq 0.46$ (none of the 800 best-fit values
implied a neutral IGM). Similarly, $88\%$ of the internal reference
neutral fraction values (distribution not shown) are in the interval
$\log(x_\mathrm{ref}) = -5.5 \pm 0.2$, and $93\%$ of the HII region
radii are within $R_\mathrm{HII} = 40.5 \pm 2 ~\mathrm{Mpc}$ (the
distributions of both of these parameters peak and have their median
at the input value). These represent the approximate confidence limits
that could be placed on these parameters from a single high-$z$ quasar
spectrum in the limit of perfect knowledge of the intrinsic quasar
spectrum, the density distribution in the IGM, and the radial profile
of the background ionizing flux. We find similar results in
simulations with $x_\mathrm{IGM} \approx 0.2$ and known intrinsic
spectra. For instance, $98\%$ of the results are at $x_\mathrm{IGM}
\leq 0.46$ in that case.

\subsection{Parameter Recovery with Extrapolated Intrinsic Spectra}

\begin{figure}
  \includegraphics[width=3in]{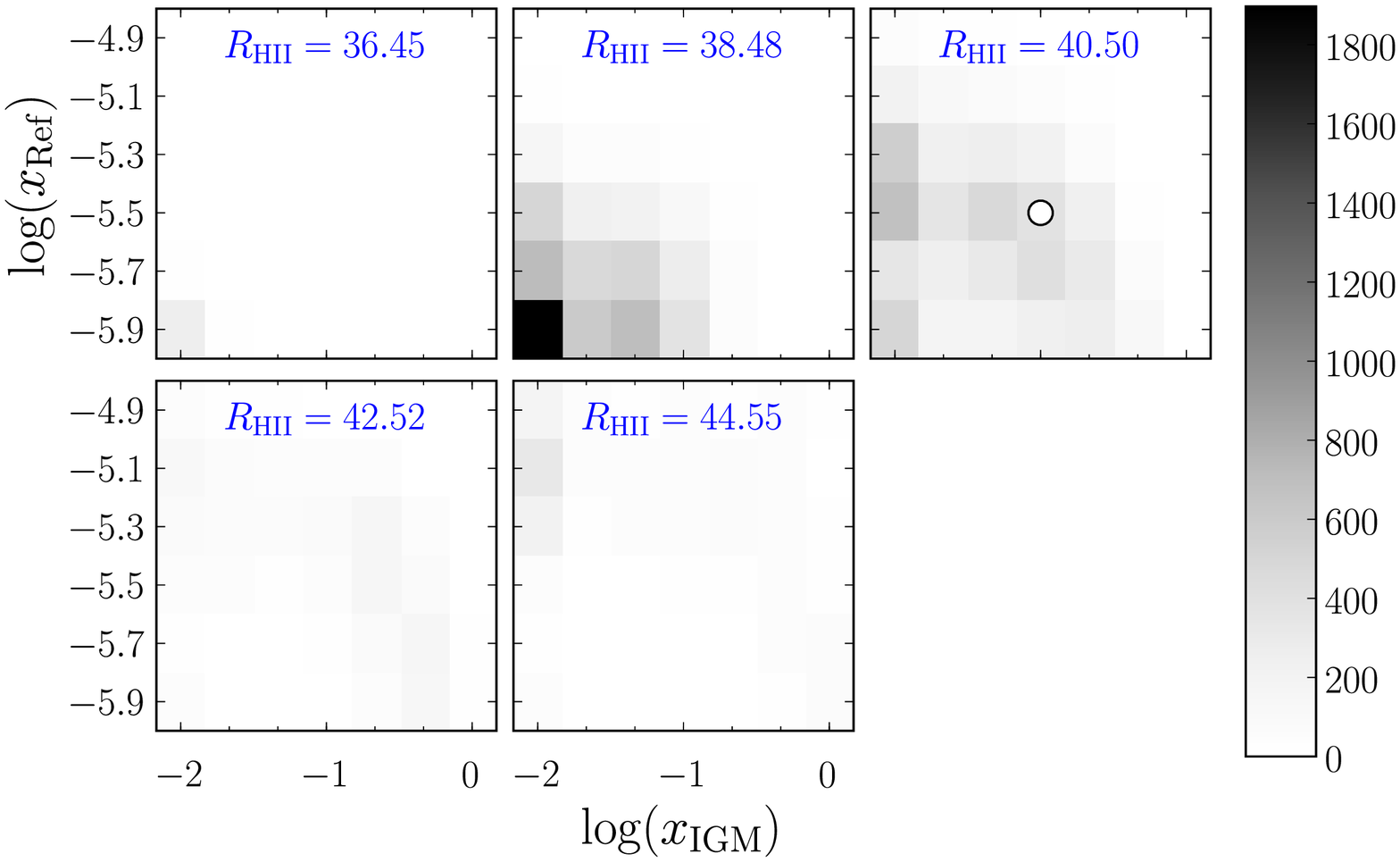}
  \includegraphics[width=3in]{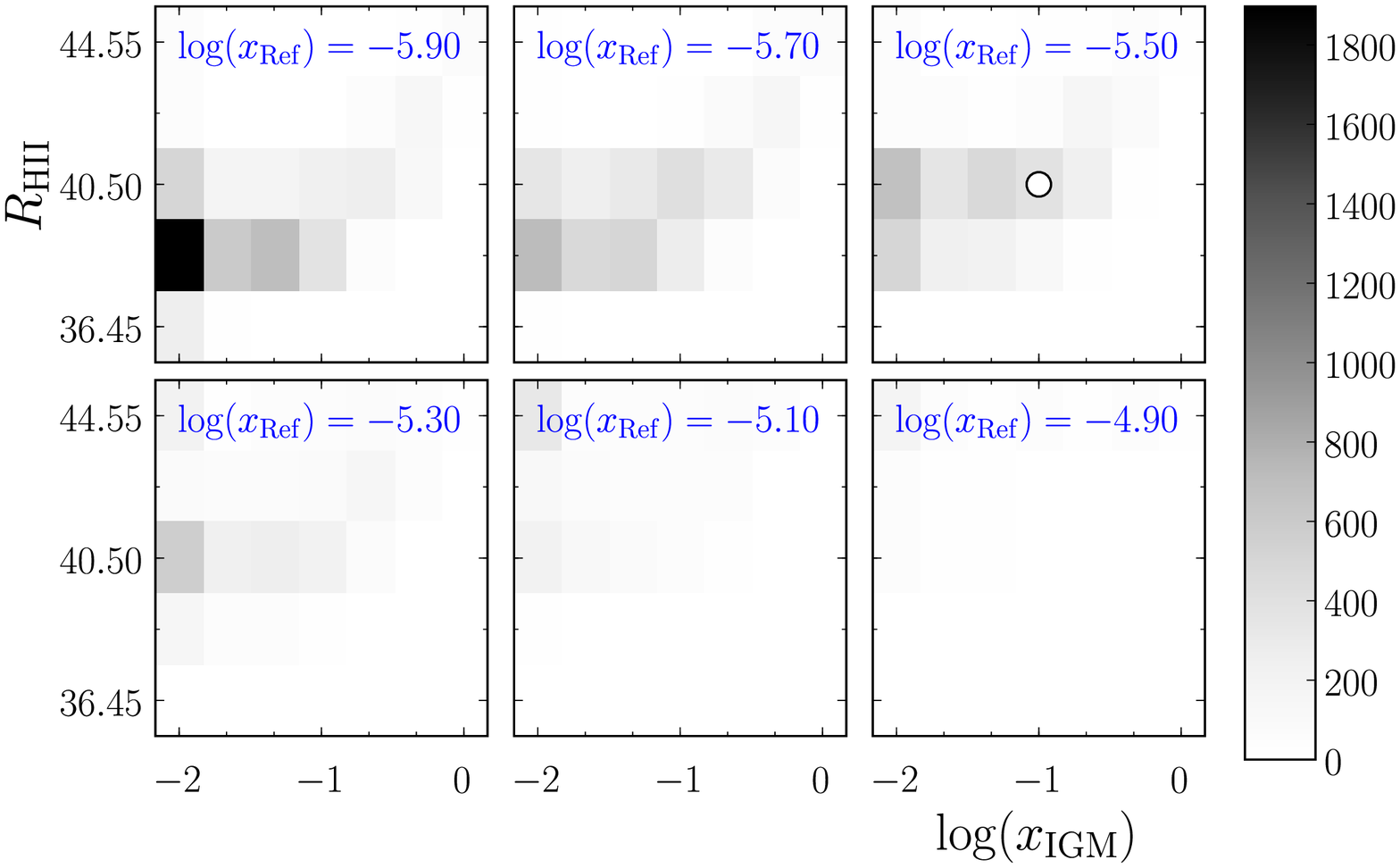}
  \includegraphics[width=3in]{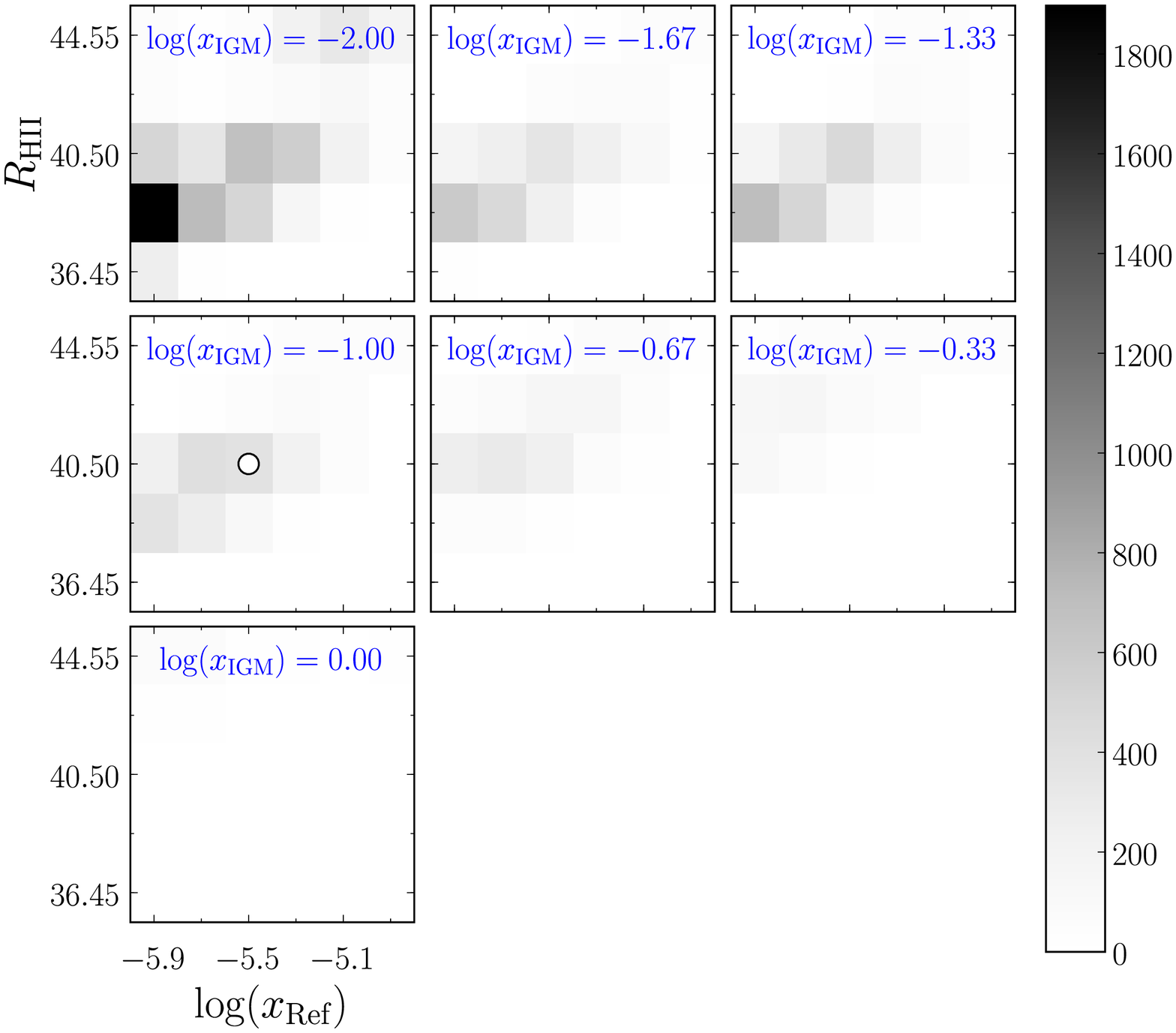}
  \caption{Parameter recovery map using mock spectra with unknown
    intrinsic spectra. The shading of each square indicates the number
    of best-fit values that fell within that cube in parameter
    space. The input parameters ($R_\mathrm{HII}=40.5\ \mathrm{Mpc}$,
    $x_\mathrm{ref} = 10^{-5.5}$, $x_\mathrm{IGM} = 0.1$) are
    indicated with a white point. Compare with Figure
    \ref{paramSpacePerfectFig}, which has the same input parameters
    but uses known intrinsic quasar spectra: an additional scatter, as
    well as a bias has been induced by errors in the extrapolation of
    the intrinsic spectrum. \label{paramSpaceFig}}
\end{figure}

In order to understand the effect that flux extrapolation errors have
on the recovered parameters, we tested IGM parameter recovery using
our more realistic mock optical depth profiles, calculated using the
spectra extrapolated from the red side of the line profile. This
mimics the process of obtaining such a profile from a high-$z$ quasar
spectrum, where the intrinsic flux is unknown. The distribution of
$17200$ best-fit values ($86$ observed spectra times $200$ random
density profiles) from the optical depth analysis using these profiles
is shown in Figure \ref{paramSpaceFig}. The same degeneracies as in
Figure \ref{paramSpacePerfectFig} (with known spectra) are evident,
but the peak of the distribution has shifted to the minimum values of
the internal and external neutral fraction, and to a lower value of
the \ion{H}{ii} region radius. Figure \ref{xIGMFig} compares the
marginalized distributions of recovered $x_\mathrm{IGM}$ values with
known and with extrapolated intrinsic spectra. Both the median and
peak of the distribution have moved to lower values, and the
``secondary'' peak at $x_\mathrm{IGM} = -2.0$, representing the
integral of the tail of the distribution that would extend to lower
values if not truncated by our finite logarithmic grid, now contains
$40\%$ of the results.

\begin{figure}
  \includegraphics[width=3in]{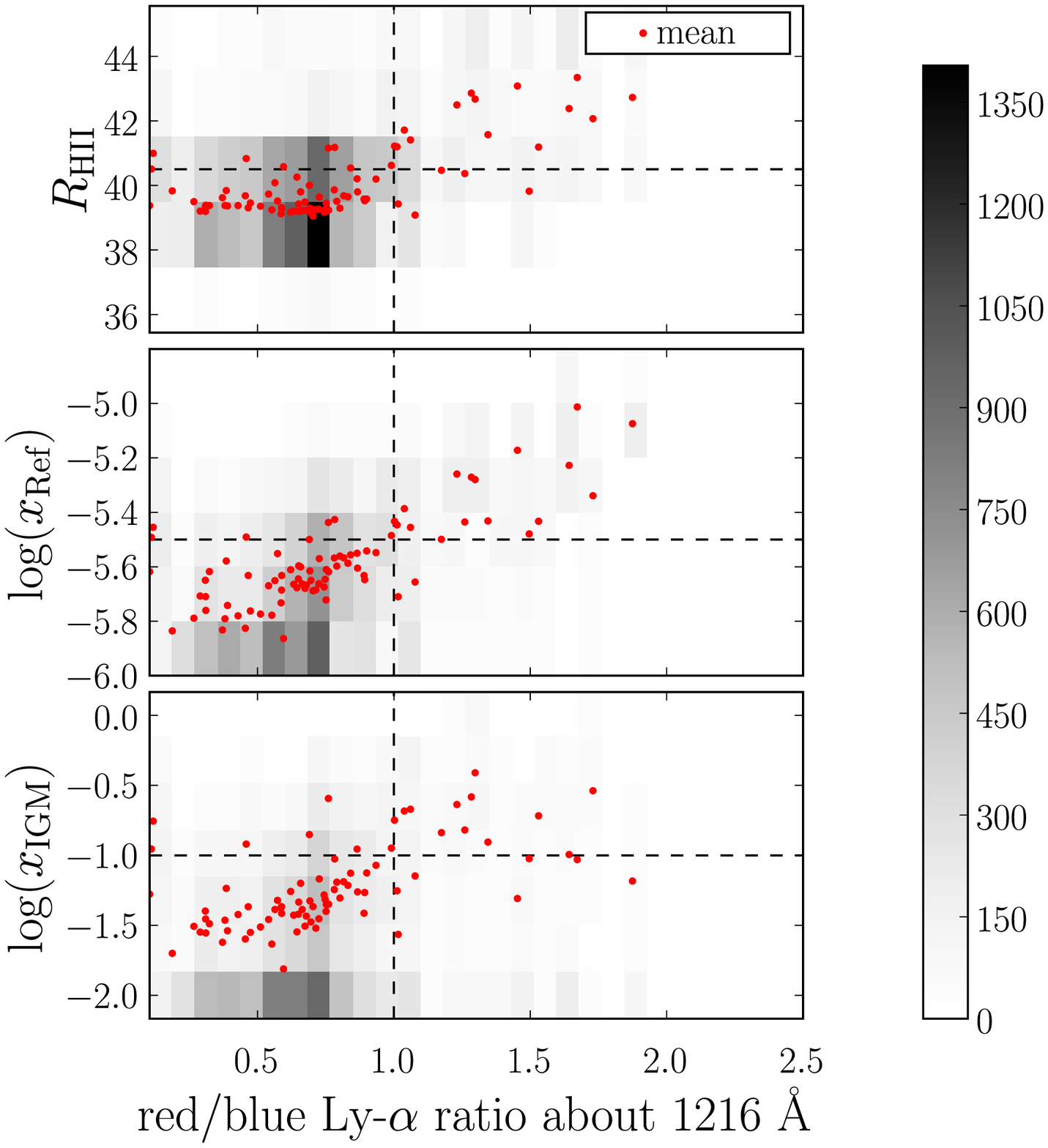}
  \caption{Best-fit parameters versus the red-to-blue flux ratio of
    the Lyman-$\upalpha$ emission about $1215.67$~\AA. The shading
    indicates how many fits fell in a given region of the graph. The
    red points plot the mean (over 200 random IGM density profiles) of the
    recovered parameter values for each input spectrum. Input
    parameters are $R_\mathrm{HII}=40.5\ \mathrm{Mpc}$,
    $x_\mathrm{ref} = 10^{-5.5}$, $x_\mathrm{IGM} = 0.1$. See
    Fig. \ref{fluxRatioFig} (top panel) for the distribution of the
    flux ratio in our sample of spectra. \label{ratioFig}}
\end{figure}

This bias in the fits toward underestimating the true neutral fraction
is caused by the bias in the flux extrapolation that we discussed in
\S \ref{fullFits}. This can be seen in Figure \ref{ratioFig}, where we
plot the recovered IGM parameters versus the red-to-blue flux ratio
about $1215.67$~\AA. The shading shows the 2D distribution of all
$17200$ fit results. The points show the mean value for each of the
$86$ intrinsic spectral models (over 200 random density
profiles). There are obvious correlations between the flux ratio and
each of the fit parameters. Because there are more spectra with excess
blue flux than excess red flux (Fig. \ref{fluxRatioFig}), and because
the red-side-only fits tend to underestimate the blue-side flux of
spectra with blue-shifted Lyman-$\upalpha$ lines (Figures
\ref{fluxvarFig} and \ref{fluxvarVsRatioFig}), there are more red-side
fits that underestimate the flux than overestimate it. If the
intrinsic flux is underestimated, then the optical depth derived from
it will also be underestimated. This tends to favor lower values of
the internal and IGM neutral fractions in the optical depth fits,
which makes sense, since both parameters correlate positively with
optical depth (though the wavelength-dependence of the bias must also
play a role in how much each parameter is affected since the
components of the optical depth have different profiles). It is a
little more surprising that the \ion{H}{ii} region radius shows the
same sign in the correlation, since it has the opposite relationship
with the optical depth. The correlation with the other parameters is
apparently strong enough to outweigh the anti-correlation with optical
depth.

So far we have tested parameter recovery with the fairly high IGM
neutral fraction of $0.1$. In light of the current constraints on the
IGM neutral fraction, however, it is interesting to explore the
question: how likely is it that an ionized IGM will be mistaken for a
neutral IGM?  More specifically, what is the probability that the best
fit values of $x_\mathrm{IGM} = 1.0, 1.0, 0.2$ for three quasars at $z
> 6.2$ would be obtained (as in \citetalias{MH2007}) if the IGM were
in fact highly ionized? To answer these questions we repeated our
analysis with the input neutral fraction lowered to $x_\mathrm{IGM} =
10^{-3}$.

\begin{figure}
  \includegraphics[width=3in]{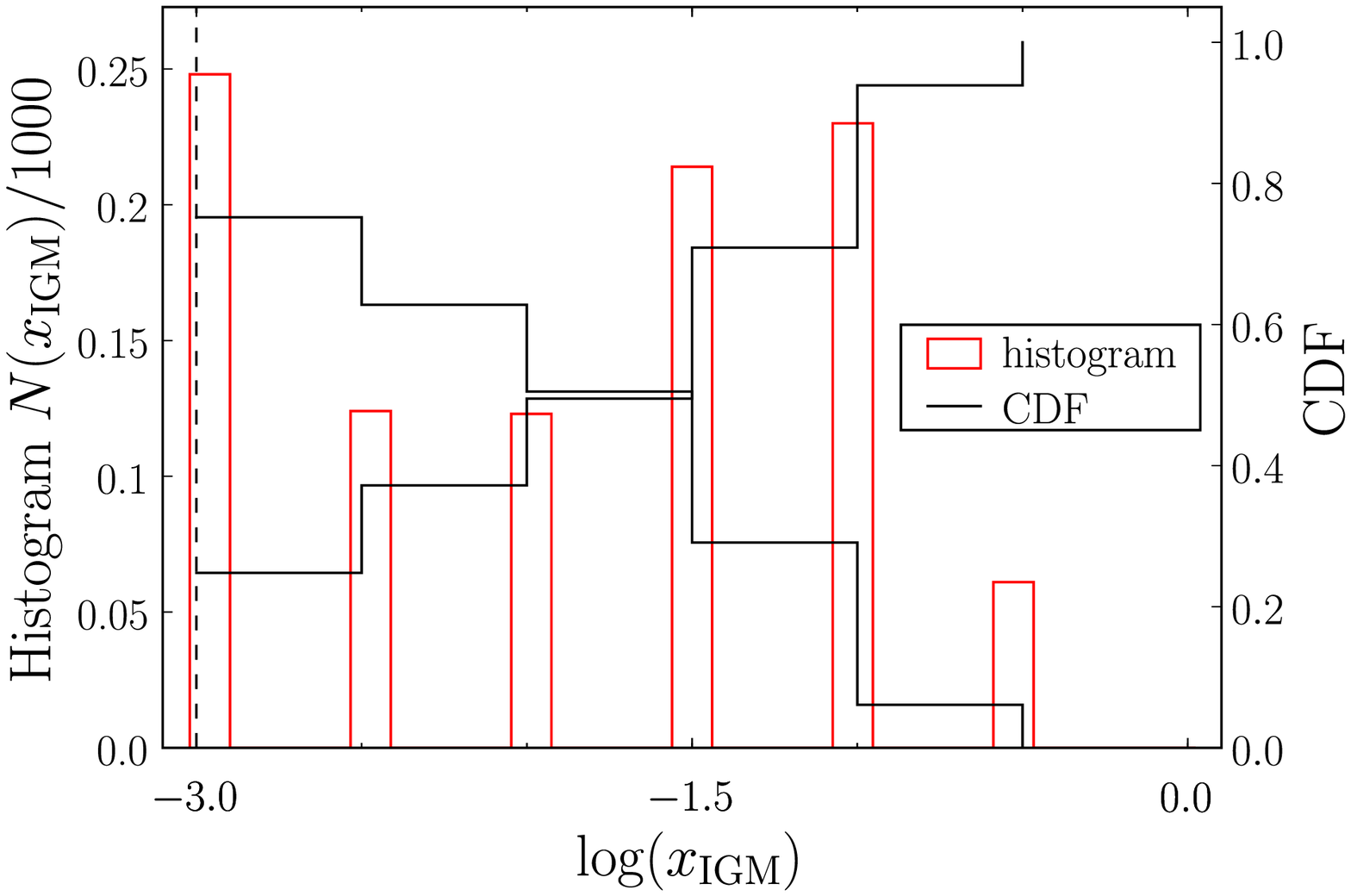}
  \includegraphics[width=3in]{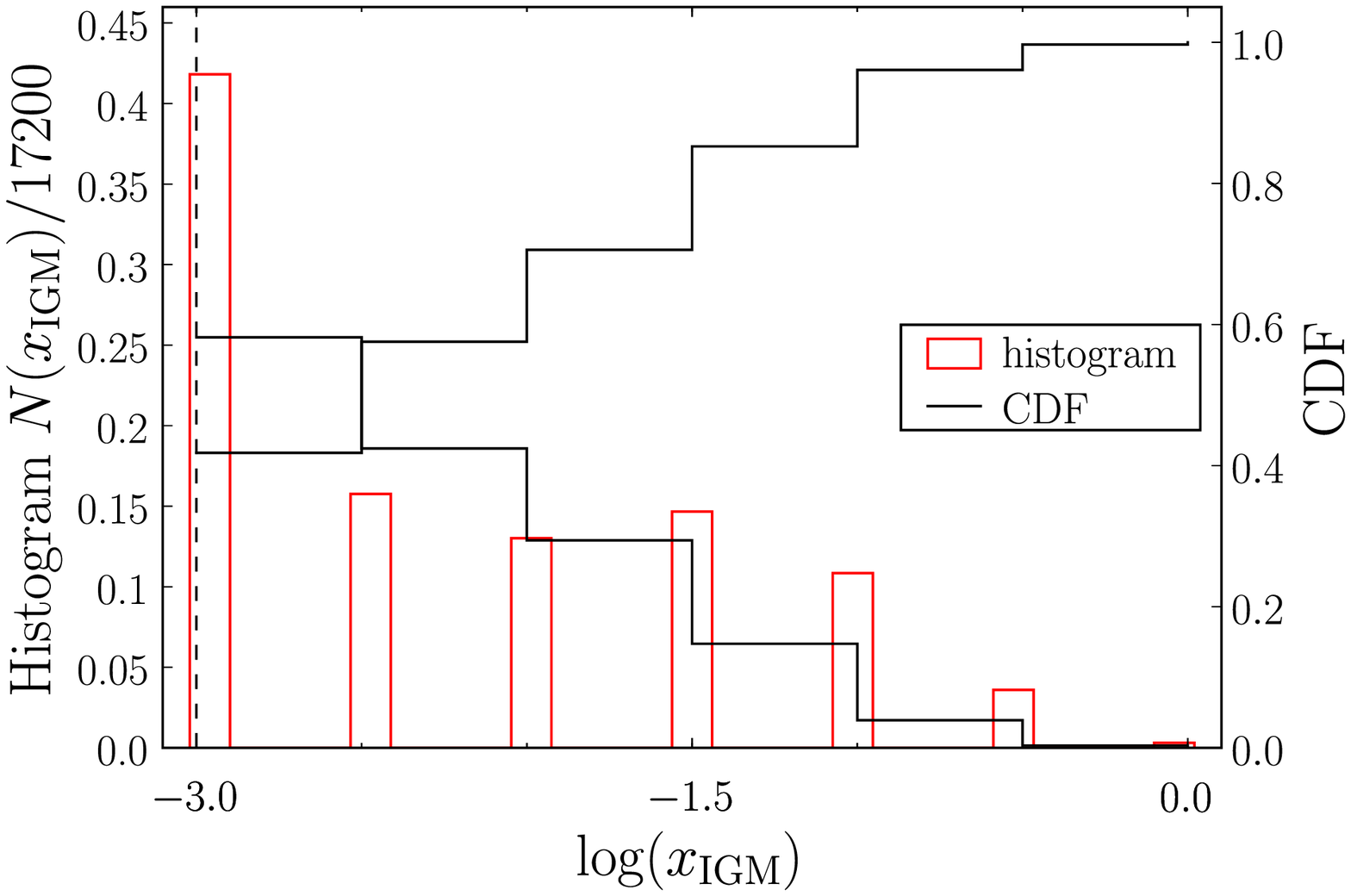}
  \caption{Marginalized (over $R_\mathrm{HII}$ and $x_\mathrm{ref}$)
    distribution of estimated $x_\mathrm{IGM}$ values with perfectly
    known (top) and extrapolated (bottom) intrinsic spectra, for a
    highly-ionized fiducial IGM. Input parameters are
    $R_\mathrm{HII}=40.5\ \mathrm{Mpc}$, $x_\mathrm{ref} = 10^{-5.5}$,
    $x_\mathrm{IGM} = 0.001$. Note that the $x$-axis scale extends to
    lower values than in Fig. \ref{xIGMFig} because we used a
    different parameter-space grid. \label{lowxIGMFig}}
\end{figure}

Figure \ref{lowxIGMFig} shows the marginalized distributions of
recovered $x_\mathrm{IGM}$ values with known and extrapolated
intrinsic spectra. Even with known intrinsic spectra the median no
longer reflects the input neutral fraction. With a highly-ionized IGM
the damping wing is too weak to be reliably detected amid the noise of
density fluctuations inside the ionized region. We can still place
strong upper limits on the neutral fraction, however. With known
intrinsic spectra (top panel), none of the best-fit values imply a
neutral IGM, $94\%$ of the recovered values were at
$\log(x_\mathrm{IGM}) \leq -0.5$, and $61\%$ at $\log(x_\mathrm{IGM})
\leq -1.0$. The added errors from using extrapolated spectra actually
strengthen the upper limit at some confidence levels due to the bias
toward underestimating the neutral fraction. $96\%$ of fits yield
$\log(x_\mathrm{IGM}) \leq -0.5$, and $85\%$ of fits yield
$\log(x_\mathrm{IGM}) \leq -1.0$. The small amount of additional
scatter to high neutral fractions is negligible. \textit{Only $0.3\%$
  of the fits yield a neutral IGM.}

\begin{figure}
  \includegraphics[width=3in]{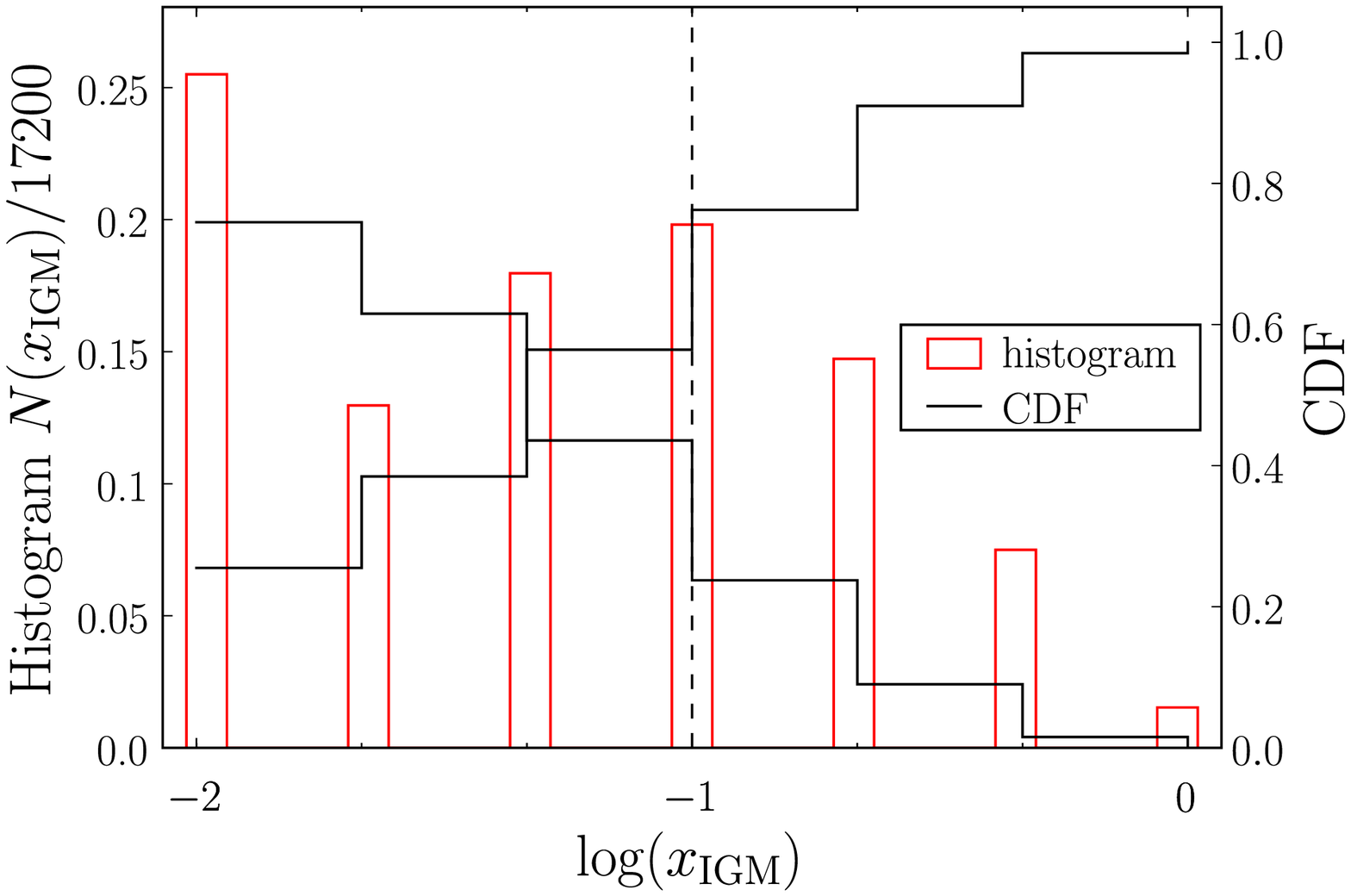}
  \caption{Marginalized recovered $x_\mathrm{IGM}$ distribution with
    the wavelength range of the fit restricted to $\lambda <
    1205$~\AA. Compare to Figure \ref{xIGMFig} (bottom panel) where
    $\lambda_\mathrm{max} = 1210$~\AA. The input parameters are
    $R_\mathrm{HII}=40.5\ \mathrm{Mpc}$, $x_\mathrm{ref} = 10^{-5.5}$,
    $x_\mathrm{IGM} = 0.1$ (as in Fig. \ref{xIGMFig}). Lowering the
    upper limit of the fit range has reduced the bias towards low
    values of $x_\mathrm{IGM}$ at the expense of some extra scatter
    toward higher values. \label{xIGMRestrictedFig}}
\end{figure}

Since the fits tend to perform worse close to the line center, we
investigated the potential to reduce the bias by lowering the maximum
wavelength of the pixels used in the optical depth fits. This cannot
be taken too far, since there are already relatively few pixels used
in these fits. Figure \ref{xIGMRestrictedFig} shows the results of
setting the maximum wavelength of our analysis region to
$\lambda_\mathrm{max} = 1205$~\AA\ rather than $1210$~\AA. The fits
(using extrapolated spectra) are dramatically improved. The bias is
reduced, with the median substantially closer to the input value and
the peak of the distribution at the input value. There is a
substantial decrease in the scatter toward lower values (only $25\%$
at $\log(x_\mathrm{IGM}) \leq -2$ versus $40\%$, and only a small
increase in the scatter toward high values. Clearly, careful choice of
the analysis region can reduce the impact of flux extrapolation
errors.

\subsection{Correcting for Median Line Shifts}\label{medianShiftSec}

\begin{figure}
  \includegraphics[width=3in]{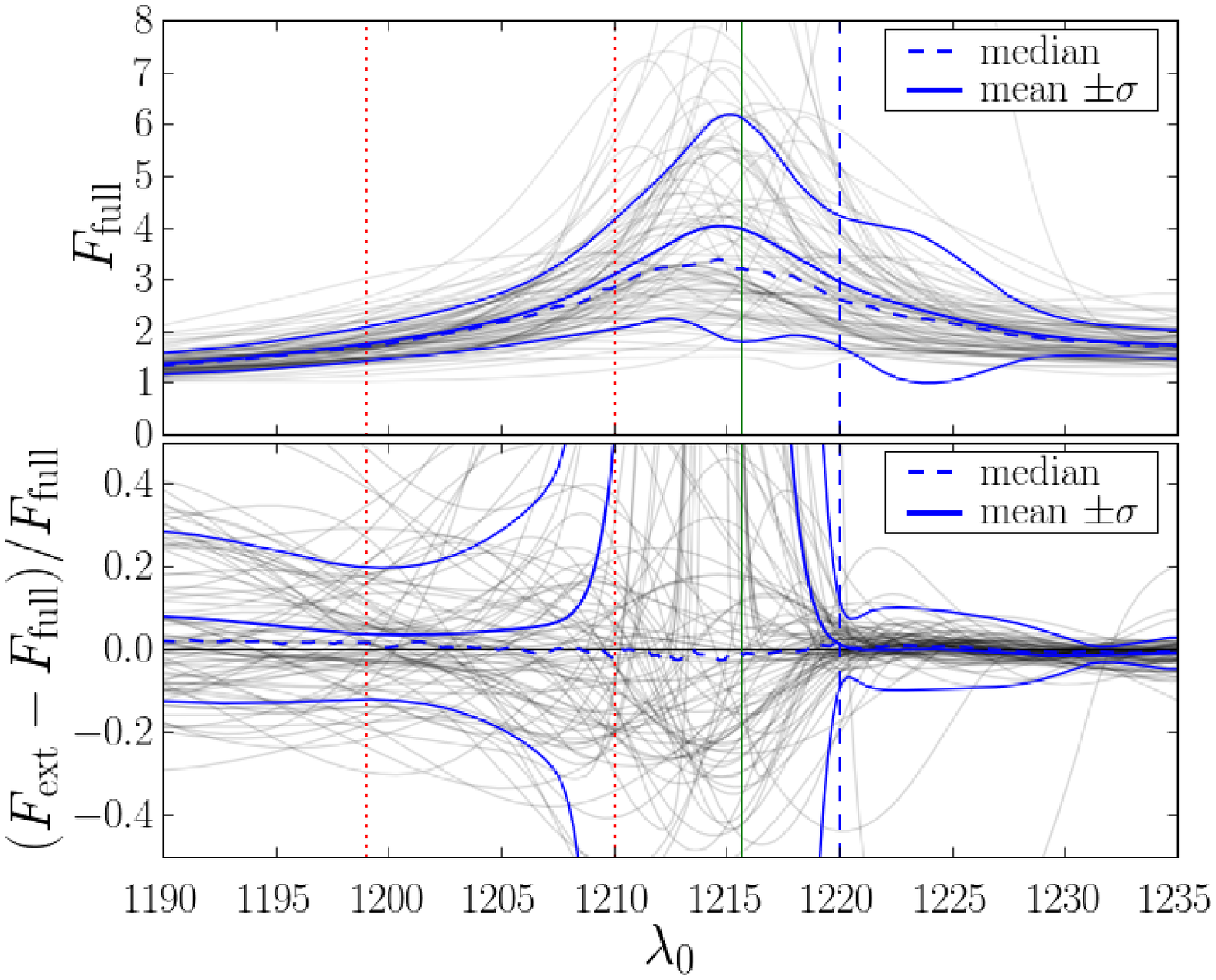}
  \caption{Fractional flux excess in the red-side-only fits (with
    velocity offsets fixed at their median values) versus wavelength. The
    excess is calculated by comparing the red-side-only best-fit model
    ($F_\mathrm{ext}$) to the full-profile fit
    ($F_\mathrm{full}$). The top panel shows the models fit to the
    full profiles of all 87 spectra (light grey) and the median
    (dashed) and mean and standard deviation (solid) of the set of
    model spectra. The bottom panel shows the fractional difference
    between the models fit to the red side only and the models fit to
    the full profile for all 87 spectra (light grey), and the median
    (dashed) and mean and standard deviation (solid) of the fractional
    flux difference. The vertical solid line (green) indicates the
    nominal central wavelength of the Lyman-$\upalpha$ line. The
    vertical dashed line (blue) at $1220$~\AA\ indicates the blue edge
    of the red-side-only line profile fit region. The vertical dotted
    lines at $1199$~\AA\ and $1210$~\AA\ (red) demarcate the typical
    analysis region for the high-redshift IGM measurements. Compare
    with Figure \ref{fluxvarshiftMedFig} (which has velocity offsets
    fixed at zero).}\label{fluxvarshiftMedFig}
\end{figure}

Since we have shown that it is the mismatch between the (on average)
blueshifted Lyman-$\upalpha$ emission and the spectral models with
line centers fixed at their nominal wavelength that is causing the
bias in the optical depth fit parameters (most importantly the IGM
neutral fraction $x_\mathrm{IGM}$), we decided to attempt to correct
for the mismatch by performing red-side-only line profile fits with
centers offset by the median shifts found in our full-profile
fits. These fits were performed as described in \S \ref{halfFits}, but
with velocity offsets fixed at $-303$, $-413$, and $-45~\mathrm{km/s}$
for the narrow Lyman-$\upalpha$, broad Lyman-$\upalpha$, and
\ion{N}{v} components. Figure \ref{fluxvarshiftMedFig} (fractional
flux excess versus wavelength) confirms that imposing a constant shift
at the median value for each component largely eliminates the bias in
the flux. As the figure shows, the median flux excess is now close to
zero everywhere. There is still a wide scatter, and the mean in fact
shows a positive bias in the core of the line. The poor performance in
the core is due to the difficulty of constraining the narrow component
with the fit range restricted to $\lambda_0 > 1220$~\AA, as discussed
in \S \ref{halfFits}.

\begin{figure}
  \includegraphics[width=3in]{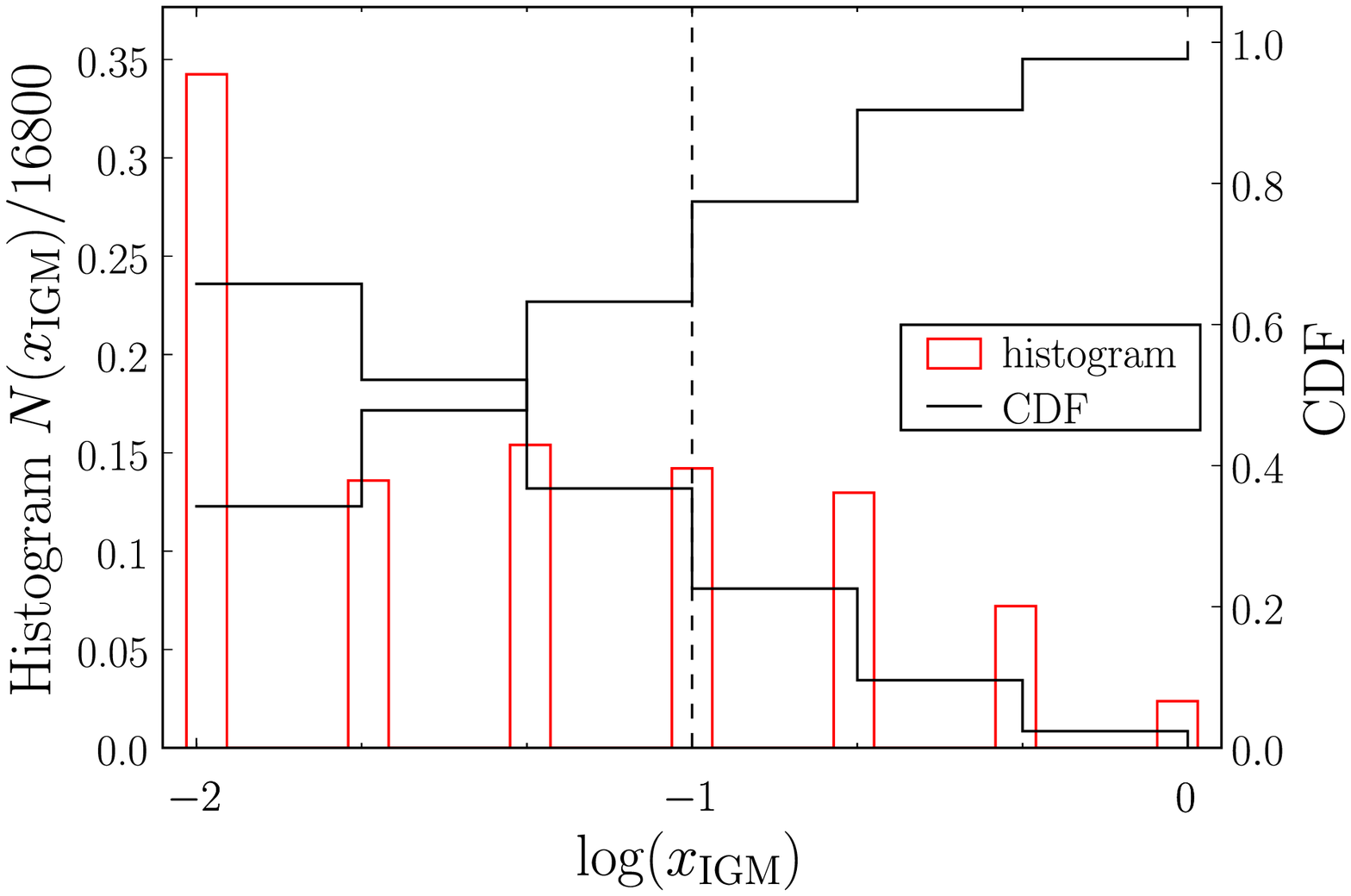}
  \caption{Marginalized recovered $x_\mathrm{IGM}$ distribution with
    velocity offsets of the components fixed at their median
    values. Compare to Figure \ref{xIGMFig} (bottom panel) where the
    offsets are fixed at zero. The input parameters are
    $R_\mathrm{HII}=40.5\ \mathrm{Mpc}$, $x_\mathrm{ref} = 10^{-5.5}$,
    $x_\mathrm{IGM} = 0.1$ (as in Fig. \ref{xIGMFig}).  Using the
    median offsets has slightly reduced the bias in the recovered
    neutral fraction distribution. Compare to Fig. \ref{xIGMFig},
    bottom panel.
    \label{xIGMshiftMedianFig}}
\end{figure}

Figure \ref{xIGMshiftMedianFig} shows the marginalized distribution of
recovered $x_\mathrm{IGM}$ values using the red-side-only fits with
median velocity offsets to extrapolate the intrinsic spectrum. The
overall distribution has shifted back toward higher values (closer to
the correct input value), correcting the bias somewhat, but the
correction is smaller than we might have expected. This crude
correction for the median offsets of the emission components is
insufficient to eliminate the bias. Interestingly, the peaks and
median values of the marginalized distributions for the \ion{H}{ii}
region radius and internal reference neutral fraction correspond to
the input values, so this simple correction has successfully
eliminated the bias for these parameters, even though it did not do so
for the IGM neutral fraction. It is unclear why this median correction
did not eliminate the neutral fraction bias. The most likely
explanation is that the neutral fraction is sensitive to the
wavelength dependence of the flux extrapolation errors in a way that
differs from the other parameters. An additional concern (discussed
earlier, in \S \ref{shiftSec}), is that the distribution of
emission-component shifts in the high-redshift quasar population may
very well differ from the distribution sampled here. This could mean
that the median values for the shifts obtained from this sample (or
other low-$z$ samples) would be biased relative to the high-$z$
population. In \S \ref{futureSec} we discuss more sophisticated
techniques for modeling and extrapolating the intrinsic spectrum that
should be more successful at eliminating the neutral fraction bias.

\subsection{Caveats}\label{conclusionSec}

While our overall conclusion that the bias is in the direction of low
$x_\mathrm{IGM}$ values should be robust, some caveats apply to the
specific numerical results discussed above:
\begin{enumerate}
\item These results assume that high-redshift quasars are drawn from
  the same population of intrinsic spectral shapes as our low-redshift
  sample. This is probably not a valid assumption in detail, since
  quasar spectra are known to show luminosity-dependent effects, but
  it may be a reasonable conservative approximation for our
  purposes. Studies of the Baldwin effect \citep{Baldwin1977,
    Espey1999, Shields2007}, indicate that the strength of the
  Lyman-$\upalpha$ line will decrease relative to the both the
  \ion{N}{v} line and the continuum in more luminous quasars. A quasar
  3 orders of magnitude more luminous than the mean for our sample
  would have a \ion{N}{v} equivalent width roughly $50\%$ larger
  assuming a Baldwin effect slope of $\sim 0.2$ \citep{Espey1999},
  though other measurements indicate a lower value for the slope
  \citep[0.0 from][]{Dietrich2002}. Similarly, the Lyman-$\upalpha$
  line would be expected to have an equivalent width roughly $30\%$
  lower for the same increase in luminosity. This means that the
  \ion{N}{v} line would be up to $2.2$ times stronger relative to the
  Lyman-$\upalpha$ line, and the \ion{N}{v} line would still
  contribute no more than a few percent of the flux in the analysis
  region. Problematic blends with other neighboring lines are
  similarly unlikely. For instance, the equivalent width of the
  \ion{Si}{ii} line at $1260$~\AA\ shows little dependence on
  luminosity \citep{Dietrich2002}, so it should still be reliably
  excluded from our fits. Direct contamination of our analysis region
  ($1199$--$1210$~\AA), would therefore be negligible, even for a
  relatively enhanced \ion{N}{v} line, but one concern is that the
  more prominent \ion{N}{v} line would in some way bias the line
  profile fits. On the other hand, a stronger \ion{N}{v} line could be
  fit to higher precision, and we have detected no tendency for the
  \ion{N}{v} line shifts to be biased, so this emission component
  might actually be modeled more accurately using high-luminosity
  quasar spectra. The fact that the Lyman-$\upalpha$ line is weaker
  relative to the continuum in luminous objects should also improve
  the accuracy of the flux extrapolation, since it is easier to
  extrapolate the continuum than the line profile, in
  general. Therefore the simultaneous \ion{N}{v} plus Lyman-$\upalpha$
  plus continuum fits may perform better at high redshift. We defer
  detailed simulation of parameter recovery with high-luminosity
  spectra to future work. Another concern is that, as we discussed in
  \S \ref{shiftSec}, indications are that a high-redshift sample will
  have larger negative velocity shifts of the Lyman-$\upalpha$ line
  relative to the metal lines. Since a larger blueward shift in the
  line center tends to result in a more severe underestimation of the
  neutral fraction, the lower limit should still be robust.

\item We assume that the mean IGM density is independent of distance
  from the quasar over our region of interest (except for the small
  change in mean density with redshift). As noted by
  \citetalias{MH2007}, the exclusion of the $6$--$11$~\AA\ region
  immediately on the blue side of the nominal line center eliminates
  the few Mpc region expected to show significant overdensity
  \citep{BarkanaLoeb2004}. However, \citet{KirkmanTytler2008},
  \citet{Guimaraes2007}, and \citet{Rollinde2005} have all inferred
  large-scale overdensities (on the scale of a few to tens of Mpc) in
  the IGM from studies of the proximity effect. The ionized regions we
  are considering are much larger ($\sim 40~\mathrm{Mpc}$). However,
  if the bias did extend into our analysis region, then it would
  increase the mean density close to the quasar. This differs from an
  error in the flux extrapolation because it shifts only the resonant
  contribution to the optical depth. Without detailed study, it is
  difficult to determine what affect this would have on the neutral
  fraction measurement. In future work, the density profile can be
  estimated from numerical cosmological simulations large enough to
  contain the rare massive host halos of bright quasars. Even if
  variations in the overdensity around such halos add to the
  uncertainty in the $x_\mathrm{IGM}$ constraint, this should not
  seriously impact the ability to distinguish highly-ionized and
  largely-neutral scenarios.

 \item We assume that the ionizing background is uniform. In reality
   the ionizing background arises from a complicated distribution of
   discrete sources, i.e. galaxies. Inside the large ionized region
   surrounding a luminous quasar, the background will still be fairly
   smooth on large scales, since the mean free path of ionizing
   photons will be large. An overdensity of galaxies would be expected
   close to the quasar, which would modify the $\tau_\mathrm{r}$ curve
   somewhat (tending to cancel out the effect of the density
   enhancement described above). If the density of galaxies is
   enhanced only very close to the quasar (on the order of a few Mpc),
   then their light will simply add to the quasar flux at larger
   radii. If, on the other hand, there is a large scale enhancement in
   the galaxy distribution extending into our analysis region ($~\sim
   10~\mathrm{Mpc}$), then the flux will not fall off like $r^{-2}$,
   and this would need to be taken into account in future analyses of
   observed quasar spectra.

 \item Outside of the quasar \ion{H}{ii} region, reionization of the
   IGM is expected to occur largely through the growth of discrete
   ionized bubbles separated by neutral gas. Our expression for the
   damping wing assumes a uniformly ionized IGM with mean neutral
   fraction $x_\mathrm{IGM}$. If, instead of being spread uniformly
   along the line of sight, the neutral gas is distributed in a
   ``picket fence'' of neutral regions separated by ionized regions,
   the damping wing will change. \citet{MF2008} studied this effect
   using semi-numerical cosmological structure simulations, concluding
   that, if ignored, this effect biases the recovered neutral fraction
   by up to $x_\mathrm{obs} - x_\mathrm{IGM} = 0.3$ (where
   $x_\mathrm{obs}$ is the value inferred from observations of the
   damping wing), and induces a scatter of a similar magnitude
   \citep[see also, ][]{McQuinn2008}. This is in the opposite
   direction to the bias we find from flux extrapolation
   errors. However, when \citet{MF2008} looked specifically at the
   bias induced by inhomogeneities around the largest halos (the
   likely hosts of bright quasars) dwelling in large ionized regions,
   they found that the bias toward overestimation of the neutral
   fraction was reduced, and even reversed for large neutral
   fractions, and that the inhomogeneity-induced modification of the
   damping wing profile may favor underestimation of the neutral
   fraction. Taking these effects together, a measured value of
   $x_\mathrm{obs} = 1$ is still unlikely if $x_\mathrm{IGM} \la
   0.02$, but more detailed simulation combined with larger quasar
   samples are needed. \citet{Lidz2007} point out another, related
   complication. The fluctuations in the ionizing background produce a
   large scatter and bias in the relationship between the apparent red
   edge of the GP trough (where the flux drops below some threshold)
   and the actual location of the ionization front. However this
   should not affect our method, since we do not assume an
   $R_\mathrm{HII}$ value in order to infer the IGM neutral fraction.

 \item We ignore radiative transfer effects, which would be especially
   important at the edge of the ionized region where the relatively
   unattenuated flux from the quasar first encounters a significant
   amount of neutral hydrogen. If the quasar spectrum is hard, the
   transition region between the highly-ionized interior and the more
   neutral exterior of the region could be broad enough to modify the
   observed optical depth profile at the edge of the transmission
   window \citep{KH2008, ThomasZaroubi2008}. It is unclear what effect
   this would have on the neutral fraction measurement.
     
\end{enumerate}

While the caveats listed above will certainly add to the scatter in
recovered values, it still seems unlikely that a highly-ionized IGM
will be mistaken for a highly-neutral one.

\section{Conclusions and Future directions}\label{futureSec}

The main conclusion of this paper is that errors in extrapolating the
intrinsic emission line shapes generally cause a bias towards
underestimating the $x_\mathrm{IGM}$ value; these uncertainties
therefore strengthen the conclusions in \citetalias{MH2007} about a
significantly neutral IGM at $z>6.2$. From the marginalized
distributions of recovered IGM parameters shown in the last section,
we can see that it is highly unlikely that flux extrapolation errors
could cause a highly-ionized IGM to be mistaken for a neutral IGM. If
$x_\mathrm{IGM} = 0.001$, there is less than a $4\%$ chance of
inferring $x_\mathrm{IGM} \ga 0.3$. Even with $x_\mathrm{IGM} = 0.1$
there is only a $15\%$ chance of measuring $x_\mathrm{IGM} \ga 0.2$
and a $5\%$ chance of measuring $x_\mathrm{IGM} \ga 0.4$. Neglecting
other sources of error, if the neutral fraction were $\leq 0.1$ there
would be less than a $0.04\%$ chance of simultaneously inferring the
three best fit values found by \citetalias{MH2007}.

Besides addressing the caveats mentioned in the last section, there
are two basic directions in which we plan to develop this technique in
the future. The first is to improve our ability to model and fit the
damping wing. This includes improving our modeling of the underlying
intrinsic quasar spectra, our models of the optical depth profiles,
and the recovery of information from the observed optical depth
profiles.

\begin{figure}
  \includegraphics[width=3in]{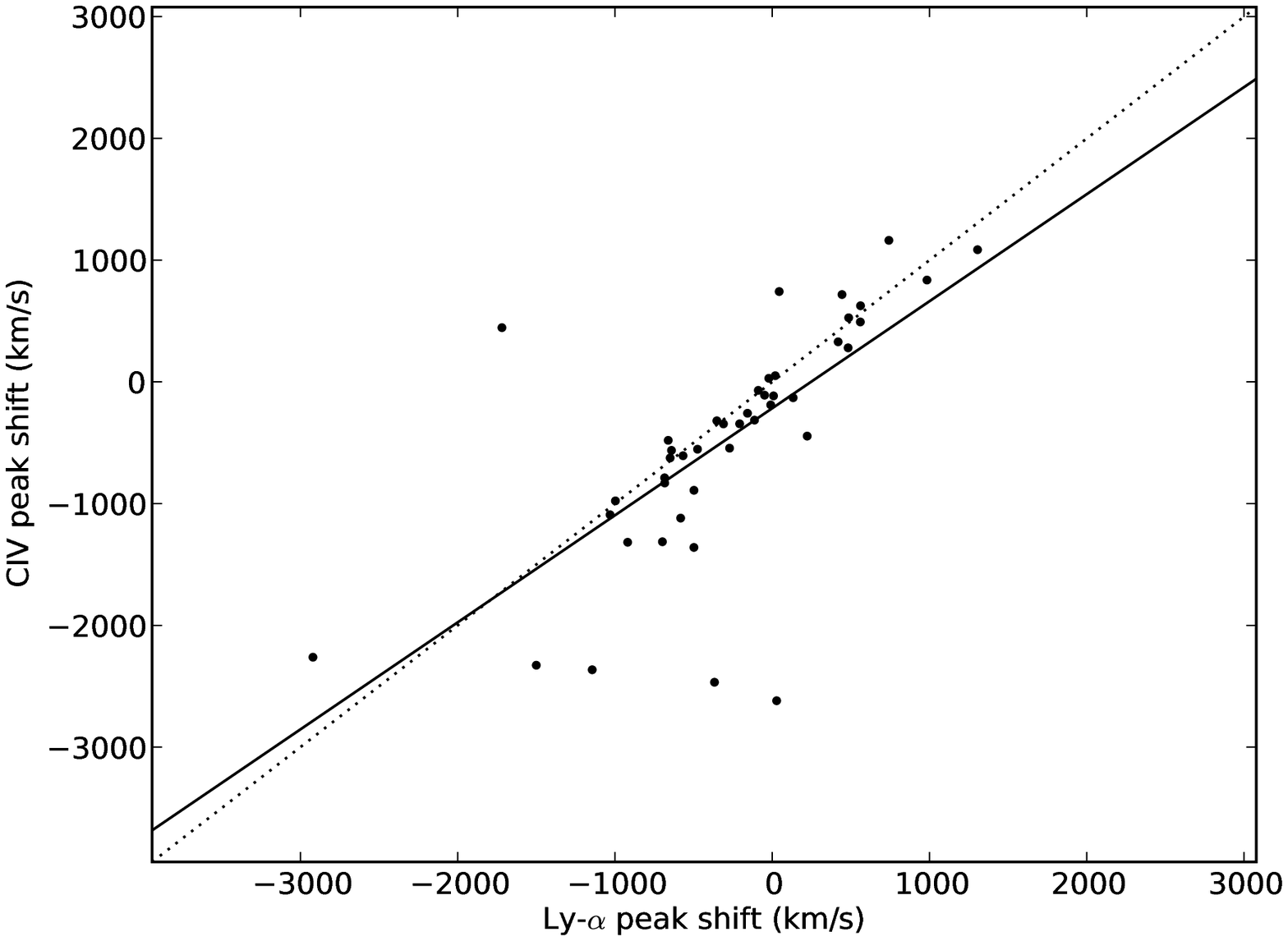}
  \caption{Shift of the \ion{C}{iv} ($1549$~\AA) line peak versus
    shift of the Lyman-$\upalpha$ peak. The solid line is the
    unweighted least-squares fit to the plotted points. The dotted
    line indicates equal shifts. Note that this plot is
    preliminary.}\label{civLyaFig}
\end{figure}

In order to improve the extrapolation of the intrinsic flux, we will
need to exploit correlations between the Lyman-$\upalpha$ emission
components and other spectral features (emission lines or continuum
luminosity), either through constraints applied to traditional
multi-component models, or through the use of principal component
analysis. For instance, \citet{Shang2007} cite a correlation between
the Lyman-$\upalpha$ and \ion{C}{iv} line shifts. We find a similar
correlation in our sample, which we show in Figure
\ref{civLyaFig}. The figure demonstrates that the shifts of the lines
(determined from fits of a quadratic function to points near the peak)
are well correlated (the correlation coefficient is $0.68$, but would
be substantially larger with the exclusion of a few
outliers). \citet{TytlerFan1992}, on the other hand, failed to find
significant correlations between the shifts of this (and any other)
pair of lines when they carefully determined systematic redshifts from
multiple emission lines. This suggests that the correlation we see may
be due, at least in part, to errors in the determination of the
systematic redshift. For our purposes, however, it does not matter why
the correlation appears, only that it could be exploited to reduce the
uncertainty on the location of the intrinsic Lyman-$\upalpha$ emission
line. In addition, one could possibly use more detailed
multi-component fits to the shape of the \ion{C}{iv} line as a
template for both the Lyman-$\upalpha$ and \ion{N}{v} lines. We expect
to be able to reduce the flux errors to the point that they are
negligible in comparison to other sources of uncertainty.

In addition to improving the spectral fits, we may be able increase
the amount of information extracted from the optical depth
profiles. Binning the optical depth values, as we have done in
performing our K-S tests, discards most of the wavelength
information. Conceivably, there is additional information in the
detailed wavelength-dependence of the optical depth profiles,
extractable with a different statistical comparison to the model
profiles.

Another source of additional information is the Lyman-$\upbeta$ region
of the absorption spectrum. While its interpretation is complicated by
the overlying Lyman-$\upalpha$ absorption, careful analysis and
comparison with simulations should allow it to supplement the
Lyman-$\upalpha$ fits \citep[e.g.][]{MH2004}.

Such improvements should be explored in order to take full advantage
of future samples of high-$z$ quasars. As surveys push deeper, more
$z>6$ quasars are constantly being discovered. For example, the
Canada-France High-z Quasar Survey \citep{Willott2009} found 10 new $z
> 5.9$ quasars, including 2 at $z > 6.2$, the SDSS Deep Stripe
\citep{Jiang2008} yielded 5 new $z > 5.85$ quasars, and UKIDSS
\citep{Mortlock2008} found one at $z = 6.13$. The quasar luminosity
function is extremely steep at the bright end that is currently being
sampled at high redshift, meaning that modest increases in the
sensitivity of future surveys should yield large increases in the
number of discovered quasars. Even if the available spectra of such
objects have significantly poorer signal-to-noise ratios than current
state-of-the-art $z > 6$ quasar spectra, they should still be suitable
for this purpose, since density fluctuations dominate the scatter in
the current results.

The techniques outlined here could also be applied to a wider range of
problems. The models (minus the IGM damping wing) and statistical
techniques described here for absorption spectra of $z>6$ quasars can
just as well be applied at $z < 6$ to probe the density and ionizing
background near quasars. As LSST and other future surveys perform deep
searches for high-$z$ objects, many lower-redshift quasars will be
discovered, which will be valuable as probes of typical quasar
environments after the end of reionization. At these lower redshifts,
rather than creating a well-defined \ion{H}{ii} region, quasars
exhibit a ``proximity effect'' \citep*{BDO1988} on the nearby
Lyman-$\upalpha$ forest, reducing the amount of absorption near the
source redshift. \citet{KirkmanTytler2008}, by studying the transverse
and line-of-sight proximity effects of quasars at $z \sim 2$, found
intriguing evidence of large-scale overdensities near the quasars, as
well as evidence for quasar lifetimes shorter than $10^6$ years. Other
previous studies (\citealp{Rollinde2005}; \citealp{Guimaraes2007};
\citealp*{Dall'Aglio2008}) have also explored the proximity effect at
$z < 4.5$. All of these groups estimated the underlying continuum by
iterative fitting to transmission windows between Lyman-$\upalpha$
forest lines. This technique is limited to $z \la 4.5$, since beyond
this redshift the Lyman-$\upalpha$ forest is too thick to reliably infer
the continuum. Above this redshift, the continuum must be extrapolated
from unabsorbed regions of the spectrum. With the improved spectral
modeling techniques we hope to develop, a uniform analysis of the
proximity effect could be performed from low redshift through $z > 6$,
without having to change the techniques used to model the underlying
spectrum.

\section{Acknowledgments}

We would like to thank Drs. Jules Halpern, Xiaohui Fan, David
Schiminovich, Greg Bryan, and David Helfand for useful discussions as
this project evolved. We would also like the thank our reviewer,
Dr. Brian Espey, for insightful and constructive feedback.

Data presented in this paper were obtained from the Multimission
Archive at the Space Telescope Science Institute (MAST). STScI is
operated by the Association of Universities for Research in Astronomy,
Inc., under NASA contract NAS5-26555. This research has made use of
the NASA/IPAC Extragalactic Database (NED) which is operated by the
Jet Propulsion Laboratory, California Institute of Technology, under
contract with the National Aeronautics and Space Administration. This
work was supported by the Pol\'anyi Program of the Hungarian National
Office for Research and Technology (NKTH).

\appendix

\section{Aligning and coadding spectra}\label{coadding}

All GHRS and some FOS (rapid-readout mode) datasets are composed of
sets of separate exposures. These separate spectra must be aligned
before being coadded to preserve spectral resolution. Also, some
objects were observed multiple times with the same instrument
configuration, and we wanted to coadd these spectra to maximize the
signal-to-noise ratio. We chose to only coadd spectra with the same
instrument configuration (instrument, grating, and aperture) so that
we would not be combining spectra with different
resolutions. Therefore we ended up with multiple spectra for some
objects.

Our code automatically tries three different methods for aligning the
spectra and chooses the one that minimizes the deviations between the
input spectra and the mean (as defined below). The methods are to
align the spectra with no offsets, based simply on the supplied
wavelength coordinates for each spectrum, to align the spectra by
maximizing the correlation function, and to first smooth the spectra,
then align them by maximizing the correlation between smoothed
spectra. The method that minimizes deviations between the aligned
spectra and the mean spectrum (as measured by $s_\mathrm{RMS}$,
defined below), is chosen for the final output. We exclude $10\%$ of
the pixels with the lowest signal-to-noise ratio in the mean spectrum,
then sum over all of the remaining pixels in each input spectrum in
order to characterize the quality of the alignment:
\begin{equation}
s_\mathrm{RMS} \equiv \sqrt{\frac{1}{n} \sum_i{ \left(\frac{f_i - \langle f \rangle_i}{E(S_i)}\right)^2}}
\end{equation}
where $f_i$ is the flux value for a pixel in an input spectrum,
$\langle f \rangle_i$ is the flux value in the corresponding pixel in
the mean spectrum, and $E(S_i)$ is the expectation value of the
standard deviation for that pixel, given by
\begin{equation}
E(S_i^2) = \frac{1}{n} \sum_j{\sigma_j^2}
\end{equation}, where
$\sigma_j$ is the uncertainty on flux value $j$, and the sum is over
all of the input pixels contributing to a single pixel in the mean
spectrum.

The general procedure for aligning and coadding spectra is as follows:

Spectra are interpolated onto a common, uniform, wavelength grid, then
(if desired) smoothed with a Gaussian kernel ($\sigma = 3$~pixels).

\textit{Aligning the spectra}: One spectrum is chosen as the reference
spectrum and the correlation function is calculated between it and
each remaining spectrum. The correlation between the reference
spectrum $F$ and another spectrum $f$ as a function of pixel offset
$s$ is 
\begin{equation}
C_{F, f}(s) = \sum_i{ \left(F_i - \langle F
  \rangle \right)  \left(f_{i+s}  - \langle
f \rangle \right) },
\end{equation}
where $F_i$ is the flux in pixel $i$ of the reference spectrum,
$f_{i+s}$ is the flux in pixel $i+s$ in the other spectrum, and
$\langle F \rangle$ and $\langle f \rangle$ are the mean flux values
of each spectrum.

For each spectrum, the offset is calculated
by finding the peak of a parabola passing through the maximum of the
correlation function and its two neighboring points. The wavelength
coordinates of the spectra are then shifted by the calculated offset
and the spectra are reinterpolated onto a common wavelength scale.

\textit{Coadding spectra}: Spectra are coadded (flux values for each
pixel are averaged), excluding outliers and choosing a weighting
scheme that maximizes the signal-to-noise. If the spectra have an
average signal-to-noise ratio of less than 2, we use an unweighted
mean of the pixel values from each spectrum. Otherwise the weights are
$w_i = F_i / \sigma_i^2$, where $F_i$ is the flux in pixel $i$ and
$\sigma_i$ is the uncertainty. If the uncertainties are
Poisson-dominated, this yields $w_i = h \nu d\nu_i dt_i A_i$ where
$d\nu$ is the bandwidth of the pixel, $dt$ is the integration time,
and $A$ is the effective area. This has the advantage over the
traditional $w_i = \sigma_i^{-2}$ of not biasing the results when the
uncertainty is estimated from the data. We also propagate the
statistical uncertainties to calculate the uncertainties for each
pixel in the weighted mean spectrum.

\textit{Excluding low signal-to-noise spectra}: If we are using the unweighted
mean, then spectra are excluded if their signal-to-noise ratio is
below $\{[n^2/(n - 1)] - (n - 1)\}^{-1/2}$ times the mean
signal-to-noise ratio. This threshold defines the approximate level
below which adding a spectrum actually degrades the overall
signal-to-noise ratio.

\textit{Normalizing spectra}: We compare the normalization of each input
spectrum to the mean by calculating the mean fractional flux offset
between it and the weighted mean spectrum. The significance of the
flux offset is then calculated, and any spectra that are more than $3
\sigma$ below the mean are normalized to have the same average flux as
the mean. We only re-normalize spectra below the mean because we
assume that flux offsets are caused by non-optimal positioning of the
source in the aperture, which can only result in missing flux. This
procedure is iterated until all mean flux differences are less than $3
\sigma$. Any bias introduced by this procedure should be unimportant,
since we only care about the overall shape of the spectrum, not its
normalization.

Outlier exclusion proceeds in two phases. First we iteratively exclude
individual pixels indentified as outliers, then we exclude all data at
wavelength coordinates where the spectrum-to-spectrum dispersion is
too high.

\textit{Excluding outlying pixel values}: We calculate the difference between
each pixel value in the input spectra and the value of that pixel in
the mean spectrum, and calculated the uncertainty of that
difference. We also calculate the standard deviation of the values for
each pixel. Any value deviating more than 5 times the propagated
uncertainty from the mean and more than 1 standard deviation from the
mean is excluded.

\textit{Flagging bad pixels}: The expected level of deviation $S$ among the $n$
data points for each pixel is given by $E(S^2) = (1/n)
\sum{\sigma_i^2}$, where $\sigma_i$ are the uncertainties on each
value.\footnote{{http://www.amstat.org/publications/jse/v13n1/vardeman.html}}
any pixel in which the standard deviation of the input values is
higher than the $3 S$ is flagged as bad and excluded from future
analysis.

Finally an output spectrum is constructed by averaging the input
spectra, with wavelength offsets and weights if appropriate, and
excluding any detected outliers.

\section{Automatic absorption feature detection}\label{featureDetection}

In order to characterize the intrinsic quasar spectra, we need to
exclude absorption features from our fits, and we need to do so
automatically to take advantage of large samples of quasars and have
reproducible results. At the same time, we want to avoid biasing our
fits by removing low-flux pixels that represent real variations in the
intrinsic quasar spectrum. We do this using two rounds of iterative
feature detection. In the first round, we apply our absorption-feature
threshold only to pixels deviating below the model spectrum. In the
second round we apply the same threshold to pixels deviating above the
model spectrum. This way, if our threshold is too stringent we will
notice spurious flagging of positive deviations. This had the
unexpected benefit of automatically detecting and excluding pixels
influenced by blending with neighboring broad emission lines in
several cases.

Our feature detection is iterative. First, we fit a model to the
observed flux profile. We then smooth the observed spectrum, model
spectrum, and the error spectra of each, and calculate the difference
between the smoothed model and observed spectra. We then calculate a
feature detection threshold
\begin{equation}
f_\mathrm{thresh}^2(\lambda) = [c_f~f_\mathrm{fit}(\lambda)]^2 +
[c_e~\sigma_\mathrm{obs}(\lambda)]^2 + \sigma_\mathrm{fit}^2(\lambda)
\end{equation}
using statistical uncertainties on the model flux
($\sigma_\mathrm{fit}$), uncertainty on the measured flux
($\sigma_\mathrm{obs}$), and the observed flux ($f_\mathrm{fit}$). The
constants $c_f = 0.05$, $c_e = 6$, were chosen by hand to be rather
conservative. That is, we err on the side of not excluding features,
rather than risk falsely excluding ``good'' pixels. Any pixel in which
the magnitude of the smoothed residuals exceeds the threshold is
flagged as an absorption or emission feature. Pixel within $\pm \Delta
\lambda$ of a flagged pixel are then excluded around the detected
features, and the fit is repeated, excluding the flagged features (or
only the flagged absorption features in the first round). 
\begin{equation}
\Delta \lambda = 2 \sqrt{ w_\mathrm{min}^2 + w_\mathrm{Inst}^2},
\end{equation}
 where $w_\mathrm{min} = 0.75$~\AA\ and $w_\mathrm{Inst} =
 \mathrm{FWHM} / 2.355$, with $\mathrm{FWHM}$ being the full width at
 half maximum of the instrumental line spread function (LSF). The
 iteration stops once no more features are flagged.

\bibliography{qso}

\end{document}